\documentclass[aps,prd,twocolumn,superscriptaddress,showpacs,showkeys,preprintnumbers]{revtex4}

\usepackage{epsfig,epsf}
\usepackage{amsmath}
\usepackage{amsthm}
\usepackage{amsfonts}
\usepackage{amssymb}
\usepackage{dsfont}

\usepackage{multirow}

\usepackage{slashed}

\usepackage[active]{srcltx}
\usepackage{psfrag}

\def\OMIT#1{}

\newcommand{\be}{\begin{equation}}
\newcommand{\ee}{\end{equation}}

\DeclareMathOperator{\Li}{Li}

\newcommand \vev [1] {\langle{#1}\rangle}

\def\II{\hbox{{1}\kern-.25em\hbox{l}}}
\begin{document}

\preprint{ \vbox{  \hbox{IPhT-t12/038}}}

\title{Finite--$t$ and target mass corrections to DVCS on a scalar target}


\date{\today}

\author{V.M.~Braun}
\affiliation{Institut f\"ur Theoretische Physik, Universit\"at
   Regensburg,D-93040 Regensburg, Germany}
\author{A.N.~Manashov}
\affiliation{Institut f\"ur Theoretische Physik, Universit\"at
   Regensburg,D-93040 Regensburg, Germany}
\affiliation{Department of Theoretical Physics,  St.-Petersburg State
University\\
199034, St.-Petersburg, Russia}
\author{B.~Pirnay}
\affiliation{Institut f\"ur Theoretische Physik, Universit\"at
   Regensburg,D-93040 Regensburg, Germany}

\date{\today}

\begin{abstract}
  \vspace*{0.3cm}
\noindent
Using the formalism developed in \cite{Braun:2011zr,Braun:2011dg}
we carry out the first complete
calculation of kinematic power corrections to the helicity amplitudes of
deeply-virtual Compton scattering to the twist-four accuracy
for a study case of a (pseudo)scalar target.
Our main result is that both finite--$t$, $\sim t/Q^2$, and
target mass, $\sim m^2/Q^2$, twist-four kinematic power corrections turn out to be
factorizable, at least to the leading order in the strong coupling. The structure of these
corrections is discussed and a short model study of their numerical impact
is presented.
 \end{abstract}



\maketitle

%
\section{Introduction}
%

It is generally accepted that studies of hard exclusive scattering processes may allow one to access a
three-dimensional picture of the proton in longitudinal and transverse plane~\cite{Burkardt:2002hr},
which is encoded in generalized parton distributions (GPDs)~\cite{Diehl:2003ny,Belitsky:2005qn}.
One of the principal reactions in this context is Compton scattering with one real and one highly-virtual photon (DVCS)
which has received a lot of attention. The corresponding measurements are planned, e.g., at
the future 12~GeV facility at the Jefferson Laboratory.

The QCD description of DVCS is based on the operator product expansion (OPE) of the time-ordered product of two electromagnetic
currents where GPDs come into play as operator matrix elements.
In order to access the transverse proton structure one is interested
in the dependence of the amplitude on the momentum transfer to the target
$t=(p'-p)^2$ in a sufficiently broad range. Since the available photon
virtualities $Q^2$ are not very large, corrections of the type $\sim t/Q^2$ which are
formally twist-four effects, can have significant impact on the data analysis
and should be taken into account. In addition, target mass corrections of the type
$\sim m^2/Q^2$ can be significant at least in certain kinematic regions and have to be included
as well.

Such effects are usually referred to as kinematic power corrections since they can be
expressed in terms of leading-twist parton distributions and do not involve ``genuine''
nonperturbative effects due to quark-gluon correlations. A well known example is provided
by Nachtmann corrections~\cite{Nachtmann:1973mr} to the structure functions in deep-inelastic
lepton-nucleon scattering (DIS).
On a technical level, these corrections arise because of the subtractions that are needed to form
traceless operators. The Nachtmann power corrections have been studied in full detail
(see e.g.~\cite{Blumlein:1998nv})  and are routinely taken into account in the analysis of DIS data.
Another classical result is due to Wandzura and Wilczek~\cite{Wandzura:1977qf}
who have shown that the twist-three structure function $g_2(x,Q^2)$ for massive targets with spin~1/2
receives a contribution related to the leading-twist structure function $g_1(x,Q^2)$. The Wandzura-Wilczek relation
is a consequence of Lorentz invariance and can be understood as spin rotation in the target rest
frame~\cite{Ball:1998sk,Anikin:2001ge}.

The importance of taking into account kinematic power corrections to DVCS
was acknowledged by many
authors~\cite{Belitsky:2005qn,Anikin:2000em,Blumlein:2000cx,Kivel:2000rb,Radyushkin:2000ap,Belitsky:2000vx,Belitsky:2001hz,Geyer:2004bx,Blumlein:2006ia,Blumlein:2008di,Belitsky:2010jw}.
This task is much more complicated, however,
because in addition to Nachtmann-type contributions related to
subtraction of traces in the leading-twist operators one must take into account
contributions of twist-four operators that are related to
the leading-twist ones by total derivatives. We call these operators descendants of the leading twist ones, 
because their matrix elements do not involve any new nonperturbative parameters. Schematically
\begin{equation}
  \mathcal{O}_1 \sim \partial^2 \mathcal{O}_{\mu_1\ldots\mu_n}\,, \qquad
  \mathcal{O}_2 \sim \partial^{\mu_1}\mathcal{O}_{\mu_1\ldots\mu_n}\,,
\label{O1O2}
\end{equation}
where $\mathcal{O}_{\mu_1\ldots\mu_n}$ are the usual leading-twist operators.
The problem arises because matrix elements of the operator $\mathcal{O}_2$ on free quarks vanish~\cite{Ferrara:1972xq}.
Thus in order to find its \emph{leading-order} coefficient function in the OPE of e.g. two electromagnetic
currents one is forced to consider either more complicated (quark-antiquark-gluon) matrix elements, or
stay with the quark-antiquark ones but go over to the next-to-leading order. In both cases the real difficulty is not the
calculation of the relevant Feynman diagrams, but the necessity to separate the contribution of interest from the contributions of
``genuine'' twist-four operators the number of which increases with the spin.
This problem was solved in Refs.~\cite{Braun:2011zr,Braun:2011dg} using hermiticity of the evolution equations for the
so-called non-quasipartonic twist-four operators with respect to a certain conformal scalar
product~\cite{Braun:2009vc}. Hermiticity implies that the coefficient functions of multiplicatively renormalizable
twist-four operators are mutually orthogonal with a proper weight function. Since ``kinematic'' twist-four
operators are multiplicatively renormalizable by construction, using this property allows one
to separate the kinematic part from an arbitrary twist-four operator in QCD ~\cite{Braun:2011dg}
and calculate the coefficient functions for all descendants of the leading twist
operators to the product of two electromagnetic currents to the required twist-four accuracy~\cite{Braun:2011zr,Braun:2011dg}.

The results of~Refs.\cite{Braun:2011zr,Braun:2011dg} amount to a complete calculation of kinematic
power corrections to twist-four accuracy in two-photon processes
at the level of the operator product expansion (to the leading order in the strong coupling).
A question remains, however, whether the OPE is applicable to the study of exclusive reactions with
one real photon, e.g. DVCS. In other language, the question is whether kinematic corrections $\sim 1/Q^2$ to DVCS 
can be taken into account consistently in the framework of collinear factorization. The answer is not obvious and, in fact, there
are many arguments suggesting that collinear factorization does not hold in DVCS beyond the leading twist.
The hope is that the collinear factorization framework may nevertheless be valid for a \emph{subset} of ``kinematic'' power
corrections defined as the contributions of the descendants of the leading-twist operators to the OPE,
because such terms are intertwined with the factorizable leading-twist contributions by the electromagnetic gauge
and, more importantly, translation invariance. In this work we verify this conjecture by  explicit
calculation on the simplest example of a (pseudo)scalar target. To this end we derive explicit expressions
for the helicity amplitudes to twist-four accuracy including all finite--$t$ and target-mass corrections,
which turn out to be remarkably simple.
An inspection shows that these amplitudes do not contain stronger singularities than those present already
in the leading-twist amplitudes and, hence, collinear factorization is not endangered
(at least to the accuracy of our calculation, i.e. in the leading order in the
strong coupling). This result is very encouraging since taking into account $\sim t/Q^2$ and $\sim m^2/Q^2$ power corrections
removes one important source of uncertainties in the theory predictions for intermediate momentum transfers,
$Q^2\sim 5-15$~GeV$^2$, which is the most interesting range in view of the planned experiments.

The presentation is organized as follows.
Sect.~2 is introductory. Here we define the kinematic variables, explain our notation and quote the necessary
portion of the results of Refs.~\cite{Braun:2011zr,Braun:2011dg} that are employed in the further analysis.
In Appendix A we explain why the results of~\cite{Braun:2011zr,Braun:2011dg} that have been derived for
flavor-nonsinglet operators are valid for flavor-singlet contributions as well, without any modification.
The calculation of helicity amplitudes for the DVCS on a scalar (or, equivalently, pseudoscalar) target is
presented in Sect.~3. In this derivation we pay special attention to the restoration of translation invariance
in the sum of all twists, leaving some details to Appendix B. The final expressions are collected in Sect.~4.
In Sect.~5 we present some model estimates of the magnitude of kinematic power corrections and summarize.

%
\section{General Formalism}
%

{}For definiteness we will consider DVCS on a pion target.
The hadronic part of the amplitude is determined by the matrix element of
the time-ordered product of two electromagnetic currents
\begin{equation}
  j^{\rm em}_\mu(x) = \bar q(x)\gamma_\mu \mathrm{Q}\, q(x)\,,
\label{jem}
\end{equation}
where $q = \{u,d\}$ is the quark field and $\mathrm{Q}$ is the diagonal matrix of quark charges
$\mathrm{Q}=e\,\text{diag}\{e_u,e_d\}$, $e=\sqrt{4\pi\alpha}$.
In the most general form
\begin{eqnarray}
\lefteqn{\hspace*{-1cm}\int\! d^4 x\int\! d^4 y\, e^{-iqx+iq'y}
\vev{\pi^b(p')|T\{j^{\rm em}_\mu(x)j^{\rm em}_\nu(y)\}|\pi^a(p)}=}
\notag\\
&=&-i (2\pi)^4\delta(p+q-p'-q')\,\mathcal{A}^{ab}_{\mu\nu}(q,q',p)\,.
\end{eqnarray}
Here $a,b$ are isospin indices, $p, p'$ are the momenta of the initial and final
state pion, and $q, q'$ are the virtual and real photon momenta, respectively:
\begin{equation}
 q'^2 = 0\,,\qquad q^2 = -Q^2\,.
\end{equation}

Making use of translation invariance one can get rid of one integration
and define the DVCS amplitude $\mathcal{A}^{ab}_{\mu\nu}$ by a simpler expression
\begin{align}\label{Az}
\mathcal{A}^{ab}_{\mu\nu}=i\!\!\int\! d^4 x\, e^{-irx}
\vev{\pi^b(p')|T\{j^{\rm em}_\mu(z_1x)j^{\rm em}_\nu(z_2x)\}|\pi^a(p)},
\end{align}
where  $z_1,z_2$ are real numbers subject to the constraint $z_1-z_2=1$ and
we use a  notation
\begin{align}\label{r}
r=z_1 q-z_2 q'\,.
\end{align}
Since $\mathcal{A}_{\mu\nu}^{ab}$ does not depend on $z_1,z_2$ one can fix their values
in some way (e.g. set $z_1=1, z_2=0$ or $z_1=-z_2=1/2$) from the very beginning.
It is instructive, however, to keep these parameters arbitrary throughout the calculation
for the following reason. In order to calculate $\mathcal{A}_{\mu\nu}^{ab}$ including terms $\sim 1/Q^2$
we have to employ the OPE  for the product of currents
\begin{align}\label{Tmn}
T_{\mu\nu}(z_1,z_2)=T\{j^{\rm em}_\mu(z_1x)j^{\rm em}_\nu(z_2x)\}\,
\end{align}
to the twist-four accuracy
\begin{align}\label{Ttwist}
T_{\mu\nu}=T^{(t=2)}_{\mu\nu}+T^{(t=3)}_{\mu\nu}+T^{(t=4)}_{\mu\nu}+\ldots\,.
\end{align}
The fact that $\mathcal{A}_{\mu\nu}^{ab}$ does not depend on $z_1, z_2$ (translation invariance) is
a consequence of the transformation property of the $T$-product:
\begin{equation}
T_{\mu\nu}(z_1+a,z_2+a) = e^{ia(\mathbf{P}x)}\,T_{\mu\nu}(z_1,z_2)\,e^{-ia(\mathbf{P}x)}\,,
\label{eq:shift1}
\end{equation}
where $\mathbf{P}$ is the usual momentum operator
\begin{equation}
   \mathbf{P}_{\mu} |p\rangle = p_\mu |p\rangle\,, \qquad i[\mathbf{P}_\mu,\Phi(x)]
  =\frac{\partial}{\partial x^\mu}\Phi(x)\,.
\label{eq:P}
\end{equation}
Equivalently,  we can write this transformation law in the differential form
\begin{equation}
\Big(\partial_{z_1}+\partial_{z_2}\Big)T_{\mu\nu}(z_1,z_2) = [(i\mathbf{P}x),T_{\mu\nu}(z_1,z_2)]\,.
\label{eq:shift2}
\end{equation}
Here and below  $[\ast\,,\ast]$ stands for a commutator.

The crucial point is that these relations do not hold for each term in the twist expansion
(\ref{Ttwist}) separately, but only in the sum of all twists. This implies that the expression for the
DVCS amplitude obtained from the contribution of the twist-two operators alone is not
translation invariant to the $O(1/Q^2)$ accuracy, i.e. it depends on the choice of the positions
of the currents. Translation invariance is restored in the sum with contributions of
higher-twist operators that are related to the leading twist by adding
total derivatives. These operators were constructed explicitly in Refs.~\cite{Belitsky:2000vx,Kivel:2000rb,Geyer:2004bx}
for twist-three and in Refs.~\cite{Braun:2011zr,Braun:2011dg} for twist-four.
This means, first, that estimates of kinematic corrections based on the Nachtmann-type contributions of
leading-twist operators alone do not have any physical significance and can be misleading.
Second, translation invariance (independence on $z_1,z_2$) of the final answer (in the sum of all twists)
provides an important check of the calculations.

A similar observation which is thoroughly discussed in the literature (see
e.g.~\cite{Anikin:2000em,Belitsky:2000vx,Kivel:2000rb}), concerns the electromagnetic gauge
invariance of the amplitude $\mathcal{A}_{\mu\nu}$, which implies the Ward identities
\begin{align}\label{qqA}
q^{\mu} \mathcal{A}_{\mu\nu}^{ab}=q'^{\nu} \mathcal{A}_{\mu\nu}^{ab}=0\,
\end{align}
or, in the operator form
\begin{eqnarray}
    \partial^\mu  T_{\mu\nu}(z_1,z_2) &=& z_2\big[ i\mathbf{P}^\mu, T_{\mu\nu}(z_1,z_2)\big]\,,
\nonumber\\
    \partial^\nu  T_{\mu\nu}(z_1,z_2) &=& z_1\big[ i\mathbf{P}^\nu, T_{\mu\nu}(z_1,z_2)\big]\,,
\label{Ward}
\end{eqnarray}
where $\partial^\mu = \partial/\partial x_\mu$.
Similar to the above, the Ward identities only hold for the sum of all twists in the OPE but are violated
for each twist separately. It was checked in Ref.~\cite{Braun:2011dg} that
Eqs.~(\ref{Ward}) are indeed satisfied in the sum of twist-two, twist-three and twist-four terms
to the required accuracy on the operator level, which guarantees that Eq.~(\ref{qqA}) is automatically
valid for scattering amplitudes from arbitrary targets. This constraint is eventually built up in the
construction of gauge-invariant, e.g. helicity, amplitudes (see below), whereas the translation invariance condition
applies to each helicity amplitude separately.

%
\subsection{Kinematics}
%
We use the two photon momenta, $q$ and $q'$, to define a longitudinal plane spanned by two
light-like vectors
\begin{equation}
n=q'\,, \qquad \tilde n=-q+(1-\tau)\, q'\,,
\end{equation}
where
\begin{equation}
  \tau= t/(Q^2+t)\,.
\end{equation}
{}For this choice the momentum transfer to the target
\begin{equation}
  \Delta = p'-p= q-q'\,,\qquad t=\Delta^2
\end{equation}
is purely longitudinal, which is convenient for calculation,
and the target (pion) momenta have a nonzero transverse component:
\begin{eqnarray}
  \Delta &=& -\tilde n-\tau n\,,
\nonumber\\
  q&=&n(1-\tau)-\tilde n\,,
\nonumber\\
P&\equiv&\frac12 (p+p')=\frac{1}{2\xi}\left(\tilde n-\tau n\right)+P_\perp\,.
\label{kinrelations}
\end{eqnarray}
The skewedness parameter $\xi$ is defined as
\begin{align}
\xi=\frac{p_+-p'_+}{p_++p'_+}\,,
\label{xi}
\end{align}
where $a_+\equiv (a n)$.
The scalar product of the two light-like vectors $n,\tilde n$ in our normalization
is of the order of $Q^2$,
\begin{align}
(n\tilde n)=\frac{Q^2}{2(1-\tau)}\,,
\end{align}
whereas
\begin{align}\label{Pperp}
| P_\perp |^2=-\left(m^2+\frac{t}{4}\frac{1-\xi^2}{\xi^2}\right)\sim \mathcal{O}(Q^0)\,.
\end{align}
Note that $t<0$ and the condition $| P_\perp |^2 > 0$ translates to the lower bound
$|t| > |t_{\rm min}| = 4m^2\xi^2/(1-\xi^2)$, cf.~\cite{Belitsky:2005qn}.

\subsection{Helicity amplitudes}

Construction of helicity amplitudes is simplified considerably in the spinor formalism.
Our conventions and notation follow Refs.~\cite{Braun:2008ia,Braun:2009vc,Braun:2011dg}.
In this approach each covariant four-vector $x_\mu$ is mapped to a hermitian $2\times2$ matrix $x_{\alpha\dot\alpha}$:
\begin{equation}
x_{\alpha\dot\alpha}=x_\mu (\sigma^{\mu})_{\alpha\dot\alpha}\,,
\qquad
\bar x^{\dot\alpha\alpha}=x_\mu (\bar\sigma^{\mu})^{\dot\alpha\alpha}\,,
\end{equation}
where $\sigma^\mu=(\II,\vec{\sigma})$, $\bar\sigma^\mu=(\II,-\vec{\sigma})$ and
$\vec{\sigma}$ are the usual Pauli matrices. We accept the following rule for raising and
lowering of spinor indices 
\begin{eqnarray}
&&u^\alpha=\epsilon^{\alpha\beta}u_\beta,\qquad u_\alpha=u^\beta\epsilon_{\beta\alpha},
\nonumber\\
&&\bar u^{\dot\alpha}=\bar u_{\dot\beta}\epsilon^{\dot\beta\dot\alpha}, \qquad
\bar u_{\dot\alpha}=\epsilon_{\dot\alpha\dot\beta}\bar u^{\dot\beta},
\end{eqnarray}
where the antisymmetric Levi-Civita tensor $\epsilon$ is defined  as follows
\begin{align}\label{eps-n}
\epsilon_{12}=\epsilon^{12}=-\epsilon_{\dot1\dot2}=-\epsilon^{\dot1\dot2}=1.
\end{align}
For this definition ${\epsilon_\alpha}^\beta=-{\epsilon^\beta}_\alpha=\delta^\beta_\alpha$
and ${\epsilon^{\dot\alpha}}_{\dot\beta}=-{\epsilon_{\dot\beta}}^{\dot\alpha}=
\delta_{\dot\beta}^{\dot\alpha}$ and
$(\epsilon^{\alpha\beta})^*=\epsilon^{\dot\beta\dot\alpha}$. An invariant product of
Weyl spinors is defined  as:
\begin{align}\label{}
(uv)=u^\alpha v_\alpha = - u_\alpha v^\alpha\,, &&
(\bar u\bar v)=\bar u_{\dot\alpha} \bar v^{\dot\alpha} = -\bar u^{\dot\alpha} \bar v_{\dot\alpha}\,.
\end{align}
More details and some useful identities can be found
in~\cite{Braun:2008ia,Braun:2009vc,Sohnius}.

An arbitrary like-like four-vector in the spinor representation, $a_{\alpha\dot\alpha}$, $a^2=0$,
can be parameterized by a Weyl spinor such that
$a_{\alpha\dot\alpha}=\xi_\alpha\bar\xi_{\dot\alpha}$, where $\bar\xi_{\dot\alpha}=\xi_{\alpha}^\dagger$.
In particular we  introduce two auxiliary spinors, $\lambda$ and $\mu$
associated with the light-like vectors $n$ and $\tilde n$ defined above:
\begin{equation}\label{ntn}
n_{\alpha\dot\alpha}=\lambda_{\alpha}\bar\lambda_{\dot\alpha}\,, \qquad
\tilde n_{\alpha\dot\alpha}=\mu_{\alpha}\bar\mu_{\dot\alpha}\,.
\end{equation}
For example one can choose
\begin{align}\label{}
\begin{pmatrix} \lambda_1\\ \lambda_2\end{pmatrix}=(E'+q'_3)^{-1/2}\begin{pmatrix} E'+q'_3\\q'_1-iq'_2\end{pmatrix}\,.
\end{align}
The scalar product $(n\tilde n)$ can be written in the form
\begin{equation}
2(n\tilde n)\equiv 2 (n_\mu \tilde n^\mu)= n_{\alpha\dot\alpha} \tilde
n^{\alpha\dot\alpha}=(\mu\lambda)(\bar\lambda\bar\mu).
\end{equation}
One of the advantages of this formalism is that two auxiliary spinors are sufficient
to define a basis in the whole four-dimensional space. Indeed, the basis vectors
in the transverse plane (orthogonal to $n,\tilde n$)
can be chosen as $\mu_{\alpha}\bar\lambda_{\dot\alpha}$ and $\lambda_{\alpha}\bar\mu_{\dot\alpha}$.
An arbitrary four-vector can be expanded in this basis as
\begin{eqnarray}
\lefteqn{\hspace*{-0.3cm}2(n\tilde n)x_{\alpha\dot\alpha}=}
\\&&{}\hspace*{-0.5cm}=
x_{++}\mu_{\alpha} \bar\mu_{\dot\alpha} + x_{--}\lambda_{\alpha}\bar\lambda_{\dot\alpha}
+x_{-+}\lambda_{\alpha}\bar\mu_{\dot\alpha}+x_{+-}\mu_{\alpha}\bar\lambda_{\dot\alpha},
\nonumber
\end{eqnarray}
where
\begin{align}
 x_{++} = \lambda^\alpha x_{\alpha\dot\alpha}\bar\lambda^{\dot\alpha}\,,&&
 x_{+-} = \lambda^\alpha x_{\alpha\dot\alpha}\bar\mu^{\dot\alpha}\,,
 \notag\\
 x_{-+} = \mu^\alpha x_{\alpha\dot\alpha} \bar\lambda^{\dot\alpha}\,,&&
 x_{--} = \mu^\alpha x_{\alpha\dot\alpha} \bar\mu^{\dot\alpha}
\end{align}
so that $x_{++}$ and $x_{--}$ correspond to the ``plus'' and ``minus'' coordinates
in the usual sense,
respectively, whereas $x_{+-}$ and $x_{-+}$ are the two (holomorphic and anti-holomorphic)
complex coordinates in the transverse plane.

Using the spinor formalism it becomes straightforward to write down a general parametrization
for the DVCS amplitude
\begin{equation}
\mathcal{A}_{\alpha\beta\dot\alpha\dot\beta}=
\sigma^{\mu}_{\alpha\dot\alpha}\sigma^{\nu}_{\beta\dot\beta} \mathcal{A}_{\mu\nu}
\end{equation}
that takes into account the Ward identities~(\ref{qqA}). The constraint
$q'^\nu \mathcal{A}_{\mu\nu}=0$ becomes $\lambda^\beta\bar\lambda^{\dot\beta}\mathcal{A}_{\alpha\beta\dot\alpha\dot\beta}=0$,
and its general solution is
\begin{align}
\mathcal{A}_{\alpha\beta\dot\alpha\dot\beta}=\lambda_\beta\bar\mu_{\dot\beta} A^{(1)}_{\alpha\dot\alpha}+
\mu_\beta\bar\lambda_{\dot\beta}A^{(2)}_{\alpha\dot\alpha}+
\lambda_\beta\bar\lambda_{\dot\beta} A^{(3)}_{\alpha\dot\alpha}\,.
\end{align}
The first two terms in this expression correspond to the contributions of a transversely polarized real photon
in the final state, and the last term to a (unphysical) longitudinally polarized photon. Making use of
the second relation in Eq.~(\ref{kinrelations}) one can easily resolve the other constraint,
$q^\mu\mathcal{A}_{\mu\nu}=0$, that results in
\begin{eqnarray}
A^{(i)}_{\alpha\dot\alpha}&=&\lambda_\alpha\bar\mu_{\dot\alpha} A^{(i,1)}+
\mu_\alpha\bar\lambda_{\dot\alpha}A^{(i,2)}
\nonumber\\&&{}
+ (\mu_\alpha\bar\mu_{\dot\alpha}+(1-\tau)\lambda_\alpha\bar\lambda_{\dot\alpha})
A^{(i,3)}\,.
\end{eqnarray}
Thus the two-photon amplitude $\mathcal{A}_{\mu\nu}$ is parameterized in the most general case by
nine scalar functions $\mathcal{A}^{(i,k)}$; three
of them correspond to a longitudinal photon in the final state and, therefore,
do not contribute to the cross-section.

In order to bring this expression to a more familiar form we define three
photon polarization vectors, $\varepsilon^{\pm,
0}_{\alpha\dot\alpha}=\sigma^{\mu}_{\alpha\dot\alpha}\varepsilon^{\pm, 0}_\mu$,
($\varepsilon^{\pm, 0}_\mu=\frac12 \sigma_{\mu}^{\alpha\dot\alpha}\varepsilon^{\pm,
0}_{\alpha\dot\alpha}$)
as follows
\begin{align}\label{}
&\varepsilon_{\alpha\dot\alpha}^+=\frac{\mu_\alpha\bar\lambda_{\dot\alpha}}{\sqrt{(n\tilde n)}}\,,
\qquad\varepsilon_{\alpha\dot\alpha}^-=\frac{\lambda_\alpha\bar\mu_{\dot\alpha}}{\sqrt{(n\tilde n)}}\,,
\notag\\
&\varepsilon_{\alpha\dot\alpha}^0=
\frac{\mu_\alpha\bar\mu_{\dot\alpha}+(1-\tau)\lambda_\alpha\bar\lambda_{\dot\alpha}}{\sqrt{2(1-\tau)(n\tilde n)}}\,.
\end{align}
The polarization vectors are normalized as
$(\varepsilon^+_\mu\varepsilon^{-\mu})=-1$, $(\varepsilon_\mu^{0})^2=1$ and 
are
orthogonal to the 
 photon momenta, $q^\mu\varepsilon^{\pm, 0}_\mu =q'^\mu\varepsilon^{\pm}_\mu=0$.
Note that $(\varepsilon^{+}_\mu)^*=\varepsilon^{-}_\mu$.

Thus we can write, finally,
\begin{align}
\!\mathcal{A}_{\mu\nu}=&\,\, \varepsilon^+_{\mu} \varepsilon^-_{\nu} \mathcal{A}^{++}
+\varepsilon^-_{\mu} \varepsilon^+_{\nu} \mathcal{A}^{--}
+\varepsilon^0_{\mu} \varepsilon^-_{\nu} \mathcal{A}^{0+}
\notag\\
&+
\varepsilon^0_{\mu}\! \varepsilon^+_{\nu} \mathcal{A}^{0-}\!
+\!\varepsilon^+_{\mu}\! \varepsilon^+_{\nu} \mathcal{A}^{+-}\!
+\!\varepsilon^-_{\mu}\! \varepsilon^-_{\nu} \mathcal{A}^{-+}\!+\!q'_\nu\mathcal{A}_\mu^{(3)}\!.
\end{align}
The amplitudes $\mathcal{A}^{\pm\pm}$   ($\mathcal{A}^{\pm\mp}$) describe virtual Compton scattering of
transversely polarized photons of positive or negative helicity without (with) a helicity flip. The
amplitudes $\mathcal{A}^{0,\pm}$ correspond to the longitudinally polarized
virtual photon in the initial state. The amplitude $\mathcal{A}^{(3)}$ is of no interest as it does
not contribute to any physical observable.

Further constraints follow from parity conservation. One obtains
\begin{eqnarray}
\mathcal{A}^{++}&=&\mathcal{A}^{--}\equiv \mathcal{A}^{(0)}\,,
\nonumber\\
\mathcal{A}^{0\pm}&=&(\varepsilon^{\pm}_\mu P^\mu)\, \mathcal{A}^{(1)}\,,
\nonumber\\
\mathcal{A}^{\mp\pm}&=&(\varepsilon^{\pm}_\mu P^\mu)^2\, \mathcal{A}^{(2)}\,.
\label{grading}
\end{eqnarray}
To explain the last two equations, note that
the light-cone vectors vectors $n$,
$\tilde n$, Eq.~(\ref{ntn}), are invariant under the redefinitions
of the auxiliary spinors by arbitrary phase factors
\begin{align}\label{lmrot}
\lambda\to e^{i\phi_1}\lambda\quad \text{and}\quad \mu\to
e^{i\phi_2}\mu\,.
\end{align}
The amplitude $\mathcal{A}_{\mu\nu}$ must stay invariant under
such rotations as well. Taking into account that with our choice of kinematics the only vector at hand which has
nonzero transverse components (and, therefore, a nonzero scalar product with the polarization vectors
$\varepsilon^{\pm}$) is $P^\mu$, one easily arrives at the representation in Eq.~(\ref{grading}).

We end up with three independent helicity amplitudes, $\mathcal{A}^{(0)}$,
$\mathcal{A}^{(1)}$, $\mathcal{A}^{(2)}$, (or, equivalently, $\mathcal{A}^{++}$,
$\mathcal{A}^{0+}$, $\mathcal{A}^{-+}$) and will calculate them in what follows.

The amplitude $\mathcal{A}_{\mu\nu}$ in vector notation can be written in terms of the helicity amplitudes in the following
way:
\begin{align}\label{covf}
\mathcal{A}_{\mu\nu}=&-g_{\mu\nu}^{\perp} \,\mathcal{A}^{(0)}+\frac{1}{\sqrt{-q^2}}\left(q_\mu-q'_\mu \frac{q^2}{(qq')}\right)
g_{\nu\rho}^{\perp} P^\rho\,\mathcal{A}^{(1)}
\notag\\
&+\frac12\left(g_{\mu\rho}^{\perp}g_{\nu\sigma}^{\perp}-\epsilon_{\mu\rho}^{\perp}\epsilon_{\nu\sigma}^{\perp}\right)
P^\rho P^\sigma\,\mathcal{A}^{(2)}+q'_\nu\mathcal{A}_\mu^{(3)}\,,
\end{align}
where
\begin{align}\label{}
g_{\mu\nu}^\perp=&g_{\mu\nu}-\frac{q_\mu q'_\nu+q'_\mu q_\nu}{(qq')}+{q'_\mu}q'_\nu\frac{q^2}{(qq')^2}\,,
\notag\\
\epsilon_{\mu\nu}^\perp=&\frac1{(qq')}\epsilon_{\mu\nu\alpha\beta}{q^\alpha q'^\beta}\,.
\end{align}
The last term $\sim q'_\nu$ in~(\ref{covf}) does not contribute to  physical observables.

\subsection{Operator product expansion}

The OPE for the product of two electromagnetic currents, Eq.~(\ref{Tmn}),
to the twist-four accuracy taking into account the leading-twist operator and its descendants obtained
by adding total derivatives is available from~\cite{Braun:2011zr,Braun:2011dg}.
In this Section we collect the necessary expressions.

To be precise, the twist-four contributions to the OPE have been obtained in~\cite{Braun:2011zr,Braun:2011dg}
for flavor-nonsinglet operators only. We are able to show, however, that the same expressions are valid (at tree level)
for flavor-singlet operators as well. This generalization is explained in Appendix A.

The tensor $T_{\alpha\beta\dot\alpha\dot\beta}(z_1,z_2)$ can be written to the twist-four accuracy
in the following form:
\begin{align}\label{Tt4}
T_{\alpha\beta\dot\alpha\dot\beta}
=&-\frac{2}{\pi^2 x^4 z_{12}^3}
\Big\{x_{\alpha\dot\beta} \mathfrak{B}_{\beta\dot\alpha}(z_1,z_2)
\notag\\
&-
x_{\beta\dot\alpha} \mathfrak{B}_{\alpha\dot\beta}(z_2,z_1)
+x_{\alpha\dot\beta} x_{\beta\dot\alpha}\, \Delta\mathbb{A}(z_1,z_2)
\notag\\
&
+x^2\Big[x_{\beta\dot\beta}\partial_{\alpha\dot\alpha}\mathbb{C}(z_1,z_2)-
x_{\alpha\dot\alpha}\partial_{\beta\dot\beta}\mathbb{C}(z_2,z_1)\Big]
\Big\}\,,
\end{align}
where $z_{12} = z_1-z_2$. (We will set $z_{12}\to 1$ in the final expressions).

The operators $\Delta\mathbb{A}$ and $\mathbb{C}$ are twist-four,
whereas the expansion of $\mathfrak{B}$ starts from the leading twist:
\begin{align}
\mathfrak{B}_{\alpha\dot\alpha}=\mathfrak{B}^{t=2}_{\alpha\dot\alpha}+
\mathfrak{B}_{\alpha\dot\alpha}^{t=3}+\mathfrak{B}^{t=4}_{\alpha\dot\alpha}+\ldots\,.
\end{align}
The twist-two result is well known and can be cast in the form
\begin{align}\label{Bt2}
\mathfrak{B}^{t=2}_{\alpha\dot\alpha}(z_1,z_2)
=\frac12\partial_{\alpha\dot\alpha}\int_0^1 du\,
\mathcal{O}_{++}^{t=2}(uz_1,uz_2)\,,
\end{align}
where $\partial_{\alpha\dot\alpha} = \sigma_{\alpha\dot\alpha}^\mu \partial_\mu$
and
\begin{align}\label{O++1}
\mathcal{O}_{++}^{t=2}(z_1,z_2)=[\Pi O_{++}](z_1,z_2)\,.
\end{align}
Here $\Pi$ is the (leading-twist) projector defined as
\begin{align}\label{Pi}
[\Pi f](x)=\sum_{k=0}^\infty\frac{(\bar\partial_{\bar\lambda}\bar x\partial_\lambda)^k}{[k!]^2}
f(\lambda_{\alpha}\bar\lambda_{\dot\alpha})|_{\lambda=0}
\end{align}
(see Ref.~\cite{Braun:2011dg} for details) and
$O_{++}(z_1,z_2)$ is a certain combination of vector and axial-vector
light-ray operators
\begin{eqnarray}\label{O++}
O_{++}(z_1,z_2)&=&
\frac12\big[O_V(z_1,z_2)-O_V(z_2,z_1)
\notag\\
&&{}-O_A(z_1,z_2)-O_A(z_2,z_1)\big]\,
\end{eqnarray}
which are defined as
\begin{eqnarray}
O_V(z_1,z_2)&=&\bar q(z_1n) \mathrm{Q}^2\,\slashed{n}q(z_2n)\,,
\notag\\
O_A(z_1,z_2)&=&\bar q(z_1n)\mathrm{Q}^2\,\slashed{n}\gamma_5 q(z_2n)\,.
\end{eqnarray}
We remind that $\mathrm{Q}$ is the matrix of quark electric  charges,
$\mathrm{Q}^2=e^2[{5}/{18}+1/6\, \tau^3]$,
cf.~(\ref{jem}). In the pion matrix element of
the operator $O_{++}$ only the vector isoscalar contribution survives: The
pion matrix element of the axial-vector operator $O_A$ vanishes identically,
whereas the matrix element of the isovector operator $O_{V}^{I=1}(z_1,z_2)$ is symmetric under
permutation of the arguments and drops out from the expression in Eq.~(\ref{O++}).
Thus for the case at hand
\begin{align}\label{ISO}
O_{++}(z_1,z_2)=\varkappa\Big[\bar u(z_1n) \slashed{n}u(z_2n)+
\bar d(z_1n) \slashed{n}d(z_2n)
\Big].
\end{align}
where 
\begin{equation}
   \varkappa = \frac{5e^2}{18}\,.
\label{varkappa}
\end{equation}
Because of this structure, the expression for scattering amplitudes from a pion are
identically the same as for a scalar target apart from a trivial Kronecker symbol in isospin indices,
$\mathcal{A}^{ab}_{\mu\nu}= \delta^{ab}\mathcal{A}_{\mu\nu}$.

The twist-three contribution to the OPE is contained in $\mathfrak{B}^{t=3}_{\alpha\dot\alpha}(z_1,z_2)$: 
\begin{widetext}
\begin{eqnarray}\label{Bt3}
\mathfrak{B}^{t=3}_{\alpha\dot\alpha}(z_1,z_2)
&=&
\!\frac14\!\!
\int_0^1\!\!\!udu\!\int_{z_2}^{z_1}\!\!\frac{dw}{z_{12}}
\biggl\{
\Big[{i\mathbf{P}}_{\!\mu},(x\bar\sigma^\mu\partial)_{\alpha\dot\alpha}z_1\mathcal{O}^{t=2}_{++}(z_1u,\! wu)
\!+\!
(\bar x\sigma^\mu\bar \partial)_{\dot\alpha\alpha}z_2\mathcal{O}^{t=2}_{++}(wu,\! z_2u)\Big]
\notag\\
&&+\ln u\,\partial_{\alpha\dot\alpha}x^2\partial_{\mu}\,
\Big[i{\mathbf{P}}^{\mu}, z_1\mathcal{O}^{t=2}_{++}(z_1u, wu)+z_2\mathcal{O}^{t=2}_{++}(wu, z_2u)\Big]
\biggr\}\,,
\end{eqnarray}
\end{widetext}
Note that the last term in this expression is by itself twist-four; its role is to subtract twist-four
contributions from the first two terms so that the result is a purely twist-three operator.

The twist-four contribution to $\mathfrak{B}_{\alpha\dot\alpha}(z_1,z_2)$ can be written as
\begin{eqnarray}\label{BB}
\mathfrak{B}^{t=4}_{\alpha\dot\alpha}(z_1,z_2)&=&x^2\partial_{\alpha\dot\alpha}\mathbb{B}(z_1,z_2)\,,
\end{eqnarray}
so that the complete twist-four contribution to the OPE is determined by three functions:
$\mathbb{B}(z_1,z_2)$, $\mathbb{C}(z_1,z_2)$ and $\Delta\mathbb{A}(z_1,z_2)=\mathbb{A}(z_1,z_2)-\mathbb{A}(z_2,z_1)$.
Explicit expressions for these functions are given in Ref.~\cite{Braun:2011dg} in several equivalent representations.
We found the following form to be the most convenient for our present purposes:
\begin{eqnarray}
\mathbb{A}(z_1,z_2)&=&\frac14\int_0^1du \,\biggl\{ u^2\ln u\,
z_1z_2 \,\mathcal{O}_1(z_1u,z_2u)
\notag\\
&&{}\hspace*{-0.9cm}+\Big(z_2\partial_{z_2}-\frac{z_1}{z_{12}}-\ln u\, z_2\partial_{z_2}^2 z_{12}\Big) \mathcal{R}(uz_1,uz_2)
\notag\\
&&{}\hspace*{-0.9cm}-\Big(z_1\partial_{z_1}-\frac{z_2}{z_{21}}-\ln u\, z_1\partial_{z_1}^2 z_{21}\Big) \bar {\mathcal{R}}(uz_1,uz_2)
\biggr\},
\nonumber\\
\label{A}
\end{eqnarray}
\begin{eqnarray}
\mathbb{B}(z_1,z_2)\!&\!=\!&\!\frac18\!\int_0^1\!\!\frac{du}{u^2} \biggl\{u^2(1\!-\!u^2\!+\!u^2\ln u)
z_1z_2\mathcal{O}_1(z_1u,z_2u)
\notag\\
&&{}\hspace*{-0.9cm}-\biggl[
(1-u^2)\left(z_2\partial_{z_2}-\frac{z_1}{z_{12}}\right)
\notag\\
&&{}\hspace*{-0.4cm}+(1\!-\!u^2\!+\!u^2\ln u)\,
z_2\partial_{z_2}^2 z_{12}\biggr] \mathcal{R}(uz_1,uz_2)
\notag\\
&&{}\hspace*{-0.9cm}+\biggl[
(1-u^2)\left(z_1\partial_{z_1}-\frac{z_2}{z_{21}}\right)
\notag\\
&&{}\hspace*{-0.4cm}+(1\!-\!u^2\!+\!u^2\ln u)\,
z_1\partial_{z_1}^2 z_{21}\biggr] \bar{\mathcal{R}}(uz_1,uz_2)\biggr\},
\label{B}
\end{eqnarray}
and
\begin{align}
\hspace*{-0.45cm}\mathbb{C}(z_1,z_2)=&-\frac{1}{8}\!\int_0^1\!\!\frac{du}{u^2}
\big[\mathcal{R}(uz_1,uz_2)+\bar{\mathcal{R}}(uz_2,uz_1)\big]\,,
\label{C}
\end{align}
where
\begin{eqnarray}\label{RbR}
\mathcal{R}(z_1,z_2)&=&
z_{12}\int_{z_2}^{z_1}\frac{d w}{z_{12}}\int_{z_2}^{w}\frac{d w'}{z_{12}} \frac{w'-z_2}{z_1-w'}
\notag\\
&&{}\hspace*{-0.9cm}\times\biggl[\frac12S_+
 \mathcal{O}_{1}(w,w' )-(S_0-1)\mathcal{O}_{2}(w,w')\biggr],
\nonumber\\
\bar{\mathcal{R}}(z_1,z_2)&=&
z_{12}
\int_{z_2}^{z_1}\frac{d w}{z_{12}}\int_{z_2}^{w}\frac{d w'}{z_{12}} \frac{z_1-w}{w-z_2}
\nonumber\\
&&{}\hspace*{-0.9cm}\times\biggl[\frac12S_+
 \mathcal{O}_{1}(w,w' )-(S_0-1)\mathcal{O}_{2}(w,w')\biggr].
\end{eqnarray}
Here  $S_+, S_0$ are differential operators acting on the light-cone coordinates
\begin{eqnarray}
S_+&=& w^2\partial_w+2w+w'^2\partial_{w'}+2w'\,,
\nonumber\\
S_0 &=&w\partial_w +w'\partial_{w'}+2\,
\end{eqnarray}
and $\mathcal{O}_{1,2}$ are nonlocal (light-ray) twist-four operators defined  in terms of the leading-twist operator
$\mathcal{O}^{t=2}_{++}(z_1,z_2)$ as follows:
\begin{align}\label{O12def}
\mathcal{O}_1(w,w')=&
\Big[i{\mathbf{P}}^{\mu},\Big[i{\mathbf{P}}_{\mu}, \mathcal{O}^{t=2}_{++}(w,w')\Big]\Big]\,,
\notag\\
\mathcal{O}_2(w,w')=&\Big[i{\mathbf{P}}^{\mu},\frac{\partial}{\partial x^\mu}
\mathcal{O}^{t=2}_{++}(w,w')\Big]\,.
\end{align}
The matrix elements of these operators are expressed in terms of the
leading-twist matrix elements (i.e. GPDs) as
\begin{eqnarray}\label{MO}
\vev{p'|\mathcal{O}_1(w,w')|p}&=&-\Delta^2\,\vev{p'|\mathcal{O}_{++}^{t=2}(w,w')|p}\,,
\nonumber\\
\vev{p'|\mathcal{O}_2(w,w')|p}&=&i\Delta^\mu\frac{\partial}{\partial x^\mu}
\vev{p'|\mathcal{O}_{++}^{t=2}(w,w')|p}\,.
\end{eqnarray}
As mentioned above, the pion matrix element of the axial-vector operator vanishes,
so that effectively $\mathcal{O}_{++}^{t=2}(z_1,z_2)$ is antisymmetric under permutation of the arguments
for the problem under consideration. As the result the operators
$\mathcal{O}_{1,2}(z_1, z_2)$ are antisymmetric in $z_1,z_2$ as well,  whereas
$\bar{\mathcal{R}}(z_1,z_2)=\mathcal{R}(z_2,z_1)$.

The reason why this representation for the twist-four contributions turns out to be
more convenient as compared to more explicit expressions in terms of $\mathcal{O}_{1,2}$
(see~\cite{Braun:2011dg}) is that the $\mathcal{R}$--operators are themselves translation-invariant
to the twist-four accuracy:
\begin{align}\label{Rshift}
\big(\partial_{z_1}+\partial_{z_2}\big)\mathcal{R}(z_1,z_2)=i\big[(\mathbf{P}x),\mathcal{R}(z_1,z_2)\big],
\end{align}
see Appendix B. This property simplifies the structure of rather intricate cancellations that
are otherwise necessary to restore translation invariance of the final results.

%
\subsection{Pion GPDs}
%

The matrix element of the operator ${O}_{++}(z_1,z_2)$ sandwiched between pions with different momenta
can be defined in terms of the (isoscalar) pion GPD~\cite{Mueller:1998fv,Ji:1996nm,Radyushkin:1996nd}
\begin{eqnarray}
\lefteqn{\hspace*{-0.6cm}\vev{\pi^b(p')|{O}_{++}(z_1n,z_2n)|\pi^{a}(p)}=}
\nonumber\\&&\hspace*{-0.6cm}=\, 2P_+\delta^{ab} \varkappa
\!\int_{-1}^1\!\!\! dx\, e^{-i P_+[z_1(\xi-x)+z_2(x+\xi)]}H(x,\xi,t)\,,
\end{eqnarray}
where $\varkappa$ is defined in Eq.~(\ref{varkappa}).

An alternative parametrization of the same matrix element is via
the double distributions (DDs)~\cite{Mueller:1998fv,Radyushkin:1997ki}.
For the isoscalar case one usually introduces two pion DDs~\cite{Polyakov:1999gs},
$f(\beta,\alpha,t)$ and $g(\beta,\alpha,t)$, so that
\begin{eqnarray}
\lefteqn{\hspace*{-0.6cm}\vev{\pi^b(p')|{O}_{++}(z_1n,z_2n)|\pi^{a}(p)}=}
\nonumber\\&=&
 i\epsilon^{abc}\varkappa \delta^{ab} \int\limits_{-1}^1 \!d\beta\!\int\limits^{1-|\beta|}_{|\beta|-1}\! d\alpha\,\,e^{-i(\ell_{z_1z_2}n)}\,
\nonumber\\&&{}\times
\Big[2(Pn)f(\beta,\alpha,t)-(\Delta n)g(\beta,\alpha,t) \Big],
\label{DD1}
\end{eqnarray}
where
\begin{equation}
\ell_{z_1 z_2}=-z_1\Delta+(z_2-z_1)\Big[\beta P-\frac12(\alpha+1) \Delta\Big].
\end{equation}
The isoscalar pion GPD is given in terms of the DDs by the following expression:
\begin{equation}
H(x,\xi,t)=\int\! d\alpha d\beta\,\delta(x-\beta-\xi\alpha)\,\Big[f(\beta,\alpha,t)+
\xi\,g(\beta,\alpha,t)\Big]\,,
\end{equation}
which can be obtained by inserting $1=\int dx\, \delta(x-\beta-\xi\alpha)$ under the integral
and comparing the definitions.

Charge conjugation and time invariance impose the following symmetry
properties of the DDs~\cite{Teryaev:2001qm}:
\begin{eqnarray}
  &&f(\beta,\alpha,t) = f(\beta,-\alpha,t)\,,\quad
  g(\beta,\alpha,t) = -g(\beta,-\alpha,t)\,,
\nonumber\\
&& f(\beta,\alpha,t) = -f(-\beta,\alpha,t)\,,\quad
  g(\beta,\alpha,t) =  g(-\beta,\alpha,t)\,.
\nonumber\\
\label{sym2}
\end{eqnarray}

The description in terms of two DDs is, however, redundant
(see a detailed discussion in~\cite{Teryaev:2001qm}), because the
expression for the matrix element is invariant under their redefinitions.
This freedom can be used to reduce $g(\beta,\alpha,t)$ to some minimal form
(the ``D-term'' \cite{Polyakov:1999gs}), but it is more convenient (see e.g.~\cite{Teryaev:2001qm,Radyushkin:2011dh}) 
to rewrite the definition in
Eq.~(\ref{DD1}) as
\begin{eqnarray}
\lefteqn{\vev{\pi^b(p')|{O}_{++}(z_1n,z_2n)|\pi^{a}(p)}=}
\nonumber\\&=&
\frac{2i}{z_{21}}\varkappa \delta^{ab}\int \! d\beta d\alpha\,
\Big[f\partial_\beta+g \partial_\alpha \Big]e^{-i(\ell_{z_1z_2}n)}
\nonumber\\&=&
\frac{2i}{z_{12}}\varkappa \delta^{ab}\int \! d\beta d\alpha\,
e^{-i(\ell_{z_1z_2}n)}\Big[\partial_\beta f + \partial_\alpha g\Big],
\label{DD2}
\end{eqnarray}
where for the last representation we assumed that $f,g$ vanish at
the boundaries.

In this form, which turns out to be the most convenient for our analysis,
the pion matrix element is parameterized in terms of a single DD
\begin{equation}\label{singleDD1}
\Phi(\beta,\alpha,t)
=\partial_\beta f(\beta,\alpha,t)+\partial_\alpha g(\beta,\alpha,t)\,,
\end{equation}
which is related to the pion GPD $H(x,\xi,t)$ as
\begin{align}
\partial_x H(x,\xi,t) =&
\int d\alpha d\beta\,\delta(x-\beta-\xi\alpha)\,\Phi(\beta,\alpha,t)\,.
\label{HPhi}
\end{align}
As a consequence of the symmetry properties (\ref{sym2}) the function
$\Phi(\beta,\alpha,t)$ is even under reflection $(\beta,\alpha)\to (-\beta,-\alpha)$, i.e.
\begin{equation}
\Phi(\beta,\alpha,t)=\Phi(-\beta,-\alpha,t)\,.
\end{equation}
We note here that the assumption that the DDs $f$ and $g$ vanish on the boundary
is made only for convenience and can be relaxed without any effect on the final result.

%
\section{Calculation of helicity amplitudes}
%

Before going into details of the calculations we want to make some general remarks.
Note that definitions of the pion GPDs involve matrix elements of nonlocal operators at
a strictly light-like separation, whereas the OPE is written in terms of operators with
arbitrary (non-light-like) separations which involve the leading-twist projectors:
\begin{eqnarray}
\lefteqn{\hspace*{-0.5cm}\vev{\pi^b(p')|\mathcal{O}^{t=2}_{++}(z_1x,z_2x)|\pi^{a}(p)}=}
\nonumber\\&=&
\Pi(x,n)
\vev{\pi^b(p')|{O}_{++}(z_1n,z_2n)|\pi^{a}(p)}\,.
\end{eqnarray}
Since the dependence of the matrix element on the quark coordinates is entirely through
the exponential factor $e^{-i(\ell_{z_1z_2}n)}$, cf. (\ref{DD2}), the leading-twist projector is
effectively applied to this exponent. It is sufficient for our purposes to know the two
first terms in the light-cone expansion~\cite{Balitsky:1987bk}
\begin{equation}\label{Px}
\Pi[e^{-i(\ell n)}](x)=e^{-i(\ell x)}+\frac{x^2\ell^2}4\int_0^1 dv\, v\, e^{-iv(\ell x)}+O(x^4)\,.
\end{equation}
Thus one obtains
\begin{eqnarray}
\vev{p'|\mathcal{O}^{t=2}_{++}(z_1x,z_2x)|p}
&=&
\frac{2i\varkappa}{z_{12}}\int \! d\beta d\alpha\,
\Phi(\beta,\alpha,t)
\nonumber\\&&\hspace*{-4.2cm}{}\times
\Big[e^{-i(\ell_{z_1z_2}n)}
+\frac{x^2\ell_{z_1z_2}^2}4\int_0^1\! dv\, v\, e^{-iv(\ell_{z_1z_2} x)}\Big].
\label{Ot=2}
\end{eqnarray}
Here and below we suppress the isospin indices.
Note that
\begin{eqnarray}
\ell_{z_1z_2}^2&=&- z_{12}^2\beta^2|P_\perp|^2 + \Delta^2\Big[z_1z_2 + z_{12}\alpha(z_{12}F-z_1)
\nonumber\\&&{}\hspace*{1cm} - z_{12}^2F(F-1)\Big]\,,
\label{ell2}
\end{eqnarray}
where we introduced a notation
\begin{align}
F=\frac12\Big(\frac{\beta}\xi+\alpha+1\Big)\,.
\label{F}
\end{align}
The second term $\sim x^2\ell_{z_1z_2}^2 $ in Eq.~(\ref{Ot=2}) involves corrections that are proportional to
the momentum transfer $t=\Delta^2$ and target mass squared $m^2$, cf.~(\ref{Pperp}).
The extra factor $x^2$ is converted to a $1/Q^2$ suppression
after the Fourier transformation (\ref{Az}), so that these contributions give rise to power suppressed,
$m^2/Q^2$ and $t/Q^2$, corrections to the helicity amplitudes. They correspond to the generalization
of the Nachtmann corrections for DIS to off-forward kinematics and have been discussed e.g.
in~\cite{Belitsky:2005qn,Belitsky:2001hz,Geyer:2004bx}.

To the same accuracy one must take into account contributions of ``kinematic'' twist-four
operators (\ref{O12def}). The leading-twist projection operator has no effect (to our accuracy)
in the case of $\mathcal{O}_1$, since the matrix element is itself proportional to $\Delta^2$,
but it must be included for $\mathcal{O}_2$ because of the
$\partial/\partial x_\mu$--derivative, see (\ref{MO}). One obtains
\begin{eqnarray}
\vev{p'|\mathcal{O}_1(z_1,z_2)|p}&=&-\frac{2i\varkappa}{z_{12}}\Delta^2\int\! d\alpha d\beta\,
\Phi
\,e^{-i(\ell_{z_1z_2},x)}\,,
\nonumber\\
\vev{p'|\mathcal{O}_2(z_1,z_2)|p}&=&\frac{2i\varkappa}{z_{12}}\int\! d\alpha d\beta\,\Phi
\Big[(\Delta \ell_{z_1z_2}) e^{-i(\ell_{z_1z_2},x)}
\nonumber\\
&&{}+\frac12(i\Delta x)\ell_{z_1z_2}^2\int_0^1\! dv v\, e^{-iv(\ell_{z_1z_2},x)}\Big],
\nonumber\\
\label{ME:O12}
\end{eqnarray}
where
\begin{equation}
  (\Delta \ell_{z_1z_2}) = -\frac12 \Delta^2[z_1+z_2-\alpha z_{12}]\,.
\end{equation}
The second term, $\sim (i\Delta x) \ell_{z_1z_2}^2$, in the expression for $\vev{p'|\mathcal{O}_2|p}$
is due to the twist projection. Note that this contribution contains terms $\sim m^2$
(through the dependence on  $|P_\perp|^2$, cf.~Eq.~(\ref{ell2})) and, therefore, the calculation of \emph{both}\ target--mass
and finite--$t$ corrections requires taking into account twist-four operators. We emphasize this point because,
naively, ``kinematic'' twist-four operators in Eq.~(\ref{O1O2}) are only relevant for the momentum transfer
dependence.

Factorization is proven for the DVCS amplitudes to the leading power accuracy only~\cite{Radyushkin:1997ki} and
there are reasons to expect that the OPE does not give a complete answer beyond the leading twist.
The main question that we address in this study is whether a subset of power corrections $m^2/Q^2$ and $t/Q^2$,
which we call kinematic, defined by the sum of contributions of the leading-twist operators and their descendants in the OPE,
is well-defined for the particular kinematics of DVCS, where one of the photons in on-shell. Our calculation will show that the
tree-level $1/Q^2$ kinematic corrections to all helicity amplitudes are finite, and their form is consistent with factorization.
Whereas this exercise does not constitute a proof, it gives a strong argument in favor of the factorization conjecture
and allows one to take into account kinematic corrections, which can be
numerically large, within the conventional framework.

In order to have the calculation under control at the intermediate steps it is convenient to
consider a more general process with two virtual photons, in which case OPE can be expected to hold to all twists,
and take the limit $(q')^2\to 0$ at the end.
Since the photon momentum $q'$ enters the calculation via the combination of momenta
$r=z_1 q-z_2 q'=q'+z_1\Delta$ only, see Eqs.~(\ref{Az}),(\ref{r}), we can redefine
\begin{equation}\label{}
   r = q'+z_1\Delta \qquad \Rightarrow \qquad r=q'+\zeta \Delta\,,
\end{equation}
and take the limit $\zeta=z_1$ at the end of calculation. This change
is equivalent to a redefinition $q'\to q'+(\zeta-z_1)\Delta$ which implies $q'^2 \slashed{=}0$ and effectively regularizes all integrals.

%
\subsection{Helicity flip amplitudes}\label{Apmop}
%
We begin with the calculation of single- and double-helicity-flip amplitudes
\begin{equation}
\mathcal{A}^{0+}=-\varepsilon_\mu^0\varepsilon_\nu^+\mathcal{A}^{\mu\nu}\,,
\qquad
 \mathcal{A}^{-+}=\varepsilon_\mu^+\varepsilon_\nu^+\mathcal{A}^{\mu\nu}\,.
\end{equation}
In chiral notations
\begin{equation}\label{Amp}
\mathcal{A}^{-+}=\frac1{4(n\tilde n)} \mathcal{A}_{--++}\equiv \frac1{4(n\tilde n)}
\mathcal{A}_{\mu\mu\bar\lambda\bar\lambda}\,
\end{equation}
and
\begin{eqnarray}\label{A0p1}
\mathcal{A}^{0+}&=&-\frac1{4(n\bar n)}\frac1{\sqrt{2(1-\tau)}}
\big[\mathcal{A}_{---+}+(1-\tau)\mathcal{A}_{+-++} \big]
\nonumber\\
&=&-\frac1{2(n\bar n)}\frac1{\sqrt{2(1-\tau)}}\mathcal{A}_{---+}\,,
\end{eqnarray}
where we have used that $\mathcal{A}_{---+}=(1-\tau)\mathcal{A}_{+-++}$ due to the
Ward identity~(\ref{qqA}).

In order to find these amplitudes one has to insert the OPE~(\ref{Tt4}) into~Eq.~(\ref{Az})
and evaluate  the corresponding integrals. Note that because of their symmetry properties
under rotations in the transverse plane the helicity-flip amplitudes must be proportional
to powers of the transverse momentum $\mathcal{A}^{0+} \propto P_\perp$,
$\mathcal{A}^{-+} \propto P_\perp^2$, multiplied
by scalar functions that depend on $\Delta$ and $P_\perp$ at least quadratically, cf. Eq.~(\ref{grading}).
If the calculation is done to the $1/Q^2$ accuracy, this dependence is irrelevant and can be neglected.
As the result, only twist-two and twist-three operators, $\mathfrak{B}^{t=2}$ (\ref{Bt2})
and $\mathfrak{B}^{t=3}$ (\ref{Bt3}), have to be taken into account
\begin{equation}
 \mathcal{A}_{---+} = \mathcal{A}^{t=2}_{---+} + \mathcal{A}^{t=3}_{---+}
\end{equation}
and similar for $\mathcal{A}_{--++}$. Twist-four operators do not contribute (to this accuracy),
which is a major simplification compared to the case of the helicity-conserving amplitude
which will be considered in the next Section.

By the same reason it is sufficient to keep the first (trivial) term
in the expansion of the leading-twist projector $\Pi$ only,~see Eq.~(\ref{Px}).
In addition, one can safely neglect all terms which contain a ``minus'' derivative:
$\partial_{--}=(\mu\partial\bar \mu)=2(\tilde n \partial)$.
Indeed, when applied to the matrix element, it produces terms
$\sim (\tilde n, \ell)\sim (\tilde n,P),\,\, (\tilde n \Delta)\sim \mathcal{O}(t,m^2)$.
Taking the matrix element of $\mathfrak{B}^{t=3}$, Eq.~(\ref{Bt3}), the action of the momentum operator $\mathbf{P}^\mu$
is effectively replaced by the multiplication by $\tilde n^\mu$:
$\mathbf{P}\to \Delta=-\tilde n-\tau n\simeq -\tilde n$.
Thus the second line in Eq.~(\ref{Bt3}) gives rise to contributions $\mathcal{O}(t,m^2)$ only and can be dropped.
It is also easy to check that only the terms proportional to $z_2$ in the first line contribute to $\mathcal{A}_{---+}$; all
other terms contain $\partial_{--}$ and can be omitted.

One finds after a short calculation
\begin{eqnarray}
\mathcal{A}_{---+}^{t=2}&=&-4\varkappa\!\int d\alpha d\beta\, \Phi(\beta,\alpha)\!\!
\int_0^1\!\! du \frac{\beta\, r_{\mu\bar\mu} P_{\mu\bar\lambda}}{(r+u\ell_{z_1z_2})^2+i\epsilon}\,,
\nonumber\\
\mathcal{A}_{---+}^{t=3}&=&\phantom{-}4\varkappa \int d\alpha d\beta\, \Phi(\beta,\alpha)\,
\int_0^1 \!duu
\notag\\
&&{}\hspace*{0.5cm}\times z_2
\int_{z_2}^{z_1}\frac{d w}{z_{12}}
\frac{\beta\, r_{\mu\bar\mu}^2P_{\mu\bar\lambda}}{((r+u\ell_{wz_2})^2+i\epsilon)^2}\,,
\end{eqnarray}
where we used that $(\mu\ell_{z_1z_2}\bar\lambda)=-\beta z_{12} P_{\mu\bar\lambda}$
since $\Delta$ does not have a transverse component.

Further, to our accuracy $r_{\mu\bar \mu}=2(n\tilde n)$ and
\begin{align}
(r+u\ell_{wz_2})^2=-2(n\tilde n)\big[z_1-uw-u(z_2-w)F\big]\,,
\end{align}
where $F$ is defined in Eq.~(\ref{F}).
Taking the integrals over $u$ and $w$ one obtains
\begin{eqnarray}
\mathcal{A}_{---+}^{t=2}&=&{4\varkappa P_{\mu\bar\lambda}}\int d\alpha d\beta \,\Phi(\beta,\alpha)\beta\,
\frac{\ln(F\!-\!i\epsilon)-\ln z_1}{F-z_1}\,,
\notag\\
\mathcal{A}_{---+}^{t=3}&=&-{4\varkappa P_{\mu\bar\lambda}}\int d\alpha d\beta \,\Phi(\beta,\alpha)\beta
\frac{z_2}{F-z_1}
\notag\\
&&{}\hspace*{1cm}\times\Big(\frac{\ln(F\!-\!i\epsilon)}{F-1}-\frac1{z_2}\ln z_1\Big)\,.
\end{eqnarray}
Note that the contributions of twist-two and twist-three operators are separately
not translation invariant, i.e. their contribution depends on $z_1,z_2$.
This dependence cancels, however, in the sum, as expected. We end up with
a very simple expression (cf.~\cite{Kivel:2000rb,Radyushkin:2000ap,Belitsky:2000vx})
\begin{align}\label{A0p}
\mathcal{A}^{0+}=\varkappa \frac{\sqrt{2} P_{\mu\bar\lambda}}{(n\tilde n)}\int d\alpha d\beta \,\Phi(\beta,\alpha)\beta\,
\frac{\ln\big(F-i\epsilon\big)}{F-1}\,,
\end{align}
which is the final result.

Calculation of the double-helicity-flip amplitude $\mathcal{A}^{-+}$ is similar.
Inserting the OPE~(\ref{Tt4}) in~Eq.~(\ref{Az})
and evaluating the Fourier integral one finds
\begin{eqnarray}
\mathcal{A}_{--++}^{t=2}\!\!&=&\!8\varkappa\!\int d\alpha d\beta\, \Phi(\beta,\alpha)
\int_0^1 \!\!du u \frac{(\beta P_{\mu\bar\lambda})^2}{(r+u\ell_{z_1z_2})^2+i\epsilon}\,,
\nonumber\\
\mathcal{A}_{--++}^{t=3}\!\!&=&\!4\varkappa r_{\mu\bar\mu}\!\int d\alpha d\beta\,
 \Phi(\beta,\alpha)(\beta P_{\mu\bar\lambda})^2
\int_0^1 \!\!du u^2\int_{z_2}^{z_1} \!\!\!dw
\nonumber\\
&&\hspace*{-1cm}\times\left[
\frac{z_1(w-z_1)}{((r\!+\!u\ell_{z_1 w})^2+i\epsilon)^2}
+
\frac{z_2(z_2-w)}{((r\!+\!u\ell_{ w z_2})^2+i\epsilon)^2}
\right].
\end{eqnarray}
After performing the $u$ and $w$ integrations the results can be written as
\begin{eqnarray}
\mathcal{A}_{--++}^{t=2}&=&-\frac{4\varkappa P_{\mu\bar\lambda}^2}{(n\tilde n)}
\int d\alpha d\beta\, \Phi(\beta,\alpha)\,\beta^2
\nonumber\\
&&\times \partial_F\left[\frac{F}{F-z_1}\ln(F-i\epsilon)-\frac{z_1}{F-z_1}\ln z_1\right],
\notag\\
\mathcal{A}_{--++}^{t=3}&=&\frac{2\varkappa P_{\mu\bar\lambda}^2}{(n\tilde n)}
\int d\alpha d\beta\, \Phi(\beta,\alpha)\,\beta^2\partial_F\biggl[\frac{\ln\big(F-i\epsilon\big)}{1-F}
\nonumber\\
&&
{}\hspace*{0.8cm}+\frac{2z_1}{F-z_1}\Big(\ln(F-i\epsilon)-\ln z_1\Big)\biggl],
\end{eqnarray}
where $\partial_F = \partial/\partial F$.
In this case, again, we observe that translation invariance
is restored in the sum of twist-two and twist-three contributions:
\begin{eqnarray}\label{Apm}
\mathcal{A}^{-+}&=&-\frac{\varkappa P_{\mu\bar\lambda}^2}{2(n\tilde n)^2}
\int d\alpha d\beta\, \Phi(\beta,\alpha)\,\beta^2
\notag\\
&&
{}\hspace*{1cm}\times\partial_F\left[\frac{2F-1}{F-1}\ln\big(F-i\epsilon\big)\right].
\end{eqnarray}
The both amplitudes, (\ref{A0p}) and (\ref{Apm}),  can be rewritten in terms
of the ``standard'' GPD $H(x,\xi,t)$.  The corresponding expressions will be
given in Sect.~\ref{results}.

%
\subsection{Helicity conserving amplitude}
%

Calculation of the helicity-conserving amplitude
\begin{equation}
   \mathcal{A}^{++} =
\varepsilon_\mu^-\varepsilon_\nu^+\mathcal{A}^{\mu\nu}
= \frac{1}{4(n\tilde n)}\mathcal{A}_{+--+}
\end{equation}
to the $\mathcal{O}(1/Q^2)$ accuracy presents our main goal.
This calculation is much more involved because of the contributions of twist-four operators.
We can write
\begin{equation}
 \mathcal{A}^{++} = \mathcal{A}^{++}_{t=2} + \mathcal{A}^{++}_{t=3} + \mathcal{A}^{++}_{t=4}\,,
\end{equation}
where $\mathcal{A}^{++}_{t=4}$ presents the main challenge.

Instead of calculating the $\mathcal{A}_{+--+}$ spinor projection of the OPE directly,
it proves to be convenient to proceed as follows.
Consider a trace of the DVCS amplitude in the Lorentz indices
\begin{equation}
{\mathcal{A}^{\mu}}_\mu=-\mathcal{A}^{++}-\mathcal{A}^{--}+\frac1{4(n\tilde n)}\mathcal{A}_{+-+-}\,.
\end{equation}
Taking into account that $\mathcal{A}^{++}=\mathcal{A}^{--}$ and
$\mathcal{A}_{-+-+}=0$ thanks  to~(\ref{qqA}), we can express the
contribution of interest as
\begin{align}\label{App}
\mathcal{A}^{++}=-\frac12{\mathcal{A}^{\mu}}_\mu+\frac1{8(n\tilde n)}
\Big[\mathcal{A}_{+-+-} - \mathcal{A}_{-+-+}\Big]\,.
\end{align}
This representation turns out to be very convenient because of strong cancellations
of different contributions in the trace and in the difference of the two terms with
different polarizations in the square brackets.

Taking the trace of the OPE in Eq.~(\ref{Tt4}) one obtains
\begin{eqnarray}
\frac12{T^{\mu}}_\mu&=&\frac1{\pi^2 x^4 z_{12}^3}\Big[
x^\mu\Delta \mathfrak{B}_\mu(z_1,z_2)+x^2\Delta \mathbb{A}(z_1,z_2)
\notag\\
&&{}-x^2(x\partial)\Delta \mathbb{C}(z_1,z_2)\Big],
\end{eqnarray}
where
\begin{eqnarray}
\Delta\mathfrak{B}_\mu(z_1,z_2)&=&\mathfrak{B}_\mu(z_1,z_2)-\mathfrak{B}_\mu(z_2,z_1)\,,
\nonumber\\
\Delta\mathbb{C}(z_1,z_2)&=&\mathbb{C}(z_1,z_2)-\mathbb{C}(z_2,z_1)\,.
\end{eqnarray}
We remind that $\mathbb{A}$ and $\mathbb{C}$ operators are of twist-four, whereas $\mathfrak{B}_\mu$
contains all twists. However, since  $x^\mu\mathfrak{B}^{t=3}_\mu(z_1,z_2)=0$ (by construction),
only $\mathfrak{B}^{t=2}$ and $\mathfrak{B}^{t=4}$ operators contribute to the trace which, therefore,
receives no twist-three contribution at all.

Making use of the explicit expression in Eq.~(\ref{Bt2})
and the identity
$$
(x\partial)\mathcal{O}^{t=2}_{++}(uz_1,uz_2)=(u\partial_u+1)\mathcal{O}_{++}(uz_1,uz_2).
$$
we can simplify the leading twist-two contribution to 
\begin{align}
x^\mu\Delta \mathfrak{B}^{t=2}_\mu(z_1,z_2)=\mathcal{O}^{t=2}_{++}(z_1,z_2)\,.
\end{align}
From now on we tacitly assume taking the matrix element over a (pseudo)scalar target in which case
$\langle p'|\mathcal{O}_{++}(z_1,z_2)|p\rangle$ is antisymmetric in $z_1,z_2$.
Similarly, for the $\mathcal{R}$--operators we use
$\mathcal{R}(z_1,z_2)=\bar{\mathcal{R}}(z_2,z_1)$ which should be understood as the
relation for the corresponding matrix elements.

The twist-four contributions to the trace can be simplified as well.
Taking into account that $x^\mu\Delta\mathfrak{B}^{t=4}_\mu(z_1,z_2)=x^2
(x\partial) \mathbb{B}(z_1,z_2)$, see Eq.~(\ref{BB}), one obtains
%
%
\begin{eqnarray}
\lefteqn{(x\partial)\Delta \mathbb{B}(z_1,z_2)\!+\!\Delta\mathbb{A}(z_1,z_2)=}
\nonumber\\
&=&
\frac14 \int_0^1\!du\, \Big[ z_1 z_2 u^2 \mathcal{O}_1(uz_1,uz_2) -
 z_2\partial_2^2z_{12}\mathcal{R}(z_1u,z_2u)
\nonumber\\&&{}\hspace*{1cm}
+z_1\partial_1^2z_{21}\mathcal{R}(z_2u,z_1u)\Big]
\end{eqnarray}
and
\begin{equation}
(x\partial)\Delta\mathbb{C}(z_1,z_2)=-\frac14\Big[\mathcal{R}(z_1,z_2)-\mathcal{R}(z_2,z_1)\Big].
\end{equation}
Next, consider the expression in the square brackets in~Eq.~(\ref{App}). In this case
only the antisymmetric in $(\alpha\dot\alpha)\leftrightarrow(\beta\dot\beta)$
part contributes to the answer
resulting in
%
\begin{eqnarray}
\lefteqn{T_{+-+-}-T_{-+-+}=}
\nonumber\\&=&
-\frac{2}{\pi^2 x^4z_{12}^3}\biggl\{x_{+-}\big[\mathfrak{B}^{t=3}_{-+}(z_1,z_2)+\mathfrak{B}^{t=3}_{-+}(z_2,z_1)\big]
\nonumber\\&&
{}\hspace*{1.4cm}-x_{-+}\big[\mathfrak{B}^{t=3}_{+-}(z_1,z_2)+\mathfrak{B}^{t=3}_{+-}(z_2,z_1)\big]
\nonumber\\&&
{}\hspace*{0.5cm}+x^2\Big[x_{--}\partial_{++}\big[\mathbb{C}(z_1,z_2)+
\mathbb{C}(z_2,z_1)\big]
\nonumber\\&&
{}\hspace*{1.05cm}-
x_{++}\partial_{--}\big[\mathbb{C}(z_1,z_2)+
\mathbb{C}(z_2,z_1)\big]\Big]\biggr\}.
\end{eqnarray}
%
The last term ($\sim x_{++}\partial_{--}$) gives rise to corrections of order $\mathcal{O}(t^2,m^2 t, m^4)$
after the
Fourier transform and can be neglected. In the remaining term $\sim x^2$ one can
replace $x_{--}\partial_{++}\to 4(n\tilde n) (x\partial) $ since the difference is again a
correction $\mathcal{O}(t^2,m^2 t, m^4)$, after which it can be combined with a similar
term $\sim \mathbb{C}$ in the trace.

Collecting everything we obtain
\begin{eqnarray}
\mathcal{A}^{++}_{t=2} &=& -\int \frac{d^4x}{\pi^2}\frac{e^{-irx}}{x^4}\langle p'|\mathcal{O}^{t=2}_{++}(z_1,z_2)|p\rangle\,,
\nonumber\\
\mathcal{A}^{++}_{t=3} &=& - \frac1{4(n\tilde n)}\int \frac{d^4x}{\pi^2}\frac{e^{-irx}}{x^4}
\nonumber\\&&{}\times\Big[
x_{+-}\langle p'|\mathfrak{B}^{t=3}_{-+}(z_1,z_2)+\mathfrak{B}^{t=3}_{-+}(z_2,z_1)|p\rangle
\nonumber\\
&&{}\hspace*{0.3cm}-x_{-+}\langle p'|\mathfrak{B}^{t=3}_{+-}(z_1,z_2)+\mathfrak{B}^{t=3}_{+-}(z_2,z_1)
|p\rangle\Big]\,,
\nonumber\\
%
\mathcal{A}^{++}_{t=4} &=&
 -\frac14\int \frac{d^4x}{\pi^2}\frac{e^{-irx}}{x^2}
\langle p'|\Big\{\!\int_0^1\!\!\!du \Big[ z_1 z_2 u^2 \mathcal{O}_1(uz_1,uz_2)
\nonumber\\&&{}
-z_2\partial_2^2z_{12}\mathcal{R}(z_1u,z_2u)
+z_1\partial_1^2z_{21}\mathcal{R}(z_2u,z_1u)\Big]
\nonumber\\&&{}
 - 2\mathcal{R}(z_2,z_1)\Big\}|p\rangle,
\label{A++t234}
\end{eqnarray}
where we suppressed  overall $1/z_{12}^3$ factors on the r.h.s. which are irrelevant
as we have to set $z_{12}\to 1$ in the final expressions.

We begin with the twist-two contribution. Using the representation for the matrix element
in Eq.~(\ref{Ot=2}) and taking the Fourier integral one obtains
\begin{eqnarray}
\mathcal{A}^{++}_{t=2} &=&
 2 \int d\alpha d\beta\,\Phi(\beta,\alpha)
\biggl\{\ln((r+\ell_{z_1z_2})^2+i\epsilon)
\nonumber\\&&{}\hspace*{0.5cm}
- \ell_{z_1z_2}^2\int_0^1 \!\!udu\,\frac{1}{(r+u\ell_{z_1z_2})^2+i\epsilon}\biggr\}.
\end{eqnarray}
Note that the second term is already $\mathcal{O}(t,m^2)$ because of $\ell_{z_1z_2}^2$, so that
we can simplify the denominator as
\begin{equation}
 (r+u\ell_{z_1z_2})^2+i\epsilon = -2(n\tilde n)[\bar u z_1+u F-i\epsilon]\,,
\end{equation}
where $\bar u =1-u$,
whereas in the first term we need an expansion of the logarithm to the $\mathcal{O}(t,m^2)$ accuracy
\begin{eqnarray}
  \ln((r+\ell_{z_1z_2})^2+i\epsilon) &=& \ln(F-i\epsilon)
\nonumber\\ &&+ \frac{1}{2(n\tilde n)}\frac{\beta^2 |P_\perp|^2}{F-i\epsilon}+\ldots.
\label{logexpand}
\end{eqnarray}
The ellipses stand for the terms that are constant or linear in $\alpha, \beta$,
which can be dropped because
\begin{equation}
\int d\alpha d\beta\,\Phi(\beta,\alpha)\, [a + b\alpha + c\beta ] = 0
\end{equation}
due to Eq.~(\ref{singleDD1}) and the reflection symmetry $\Phi(\beta,\alpha) = \Phi(-\beta,-\alpha)$.
Using
\begin{equation}
\int_0^1 \!\frac{udu}{\bar u z_1+u F-i\epsilon} = \partial_F
\Big[\frac{F\ln (F-i\epsilon)}{F-z_1}-\frac{z_1\ln z_1}{F-z_1}\Big]
\end{equation}
we obtain
\begin{eqnarray}
\mathcal{A}^{++}_{t=2} &=&
  {2}\int d\alpha d\beta\,\Phi(\beta,\alpha)
\biggl\{
\ln(F-i0)
\nonumber\\&&{}
+ \frac{1}{2(n\tilde n)}\biggl\{-\beta^2 |P_\perp|^2\partial_F\frac{z_1}{F-z_1}\ln \Big(\frac{F}{z_1}-i\epsilon\Big)
\nonumber\\&&{}+
\Delta^2\big[z_1z_2 + (F-z_1)\alpha- F\left(F-1\right)\big]
\nonumber\\&&{}\hspace*{0.5cm}\times
\partial_F \Big[\frac{F\ln (F-i\epsilon)}{F-z_1}-\frac{z_1\ln z_1}{F-z_1}\Big]
\biggr\}.
\label{A++t=2}
\end{eqnarray}

The calculation of the twist-three contribution follows the scheme described
in Sect.~(\ref{Apmop}). We mention only that the second line in Eq.~(\ref{Bt3})
does not contribute to the result due to the antisymmetry in $z_1,z_2$. Also, in this calculation
it is sufficient to take into account the leading term in the twist projection operator~(\ref{Px}) only.
A straightforward calculation yields
\begin{eqnarray}
\mathcal{A}^{++}_{t=3}&=&
-\frac1{(n\bar n)}\int d\alpha d\beta\, \Phi(\beta,\alpha)\, \beta\,
\nonumber\\&&{}\times
\Big[\frac{\Delta^2}\xi-\beta|P_\perp|^2\partial_F\Big]\frac{\ln(F-i\epsilon)}{F-1}\,,
\label{A++t=3}
\end{eqnarray}
where $|P_\perp|^2$ is defined in Eq.~(\ref{Pperp}).

Calculation of the twist-four contribution is somewhat longer.
It is given by a sum of four terms,
\begin{equation}
 \mathcal{A}^{++}_{t=4} =  \mathcal{A}^{(1)}_{t=4} + \mathcal{A}^{(2)}_{t=4} +\mathcal{A}^{(3)}_{t=4} +\mathcal{A}^{(4)}_{t=4}
\end{equation}
which we numerate in the same order as they appear in the last equation in~(\ref{A++t234}).

The first contribution, of the operator $\mathcal{O}_1$, is really simple.
A short calculation yields
\begin{eqnarray}
 \mathcal{A}^{(1)}_{t=4} &=& -\frac{z_1z_2\Delta^2}{(n\tilde n)}
\!\int\! d\alpha d\beta\,\Phi(\beta,\alpha)
\nonumber\\&&{}\times
\partial_F
\Big[
  \frac{F \ln (F-i\epsilon)}{F-z_1}-\frac{z_1\ln z_1}{F-z_1}
\Big].
\end{eqnarray}

The remaining three contributions involve the $\mathcal{R}$--operator defined in Eq.~(\ref{RbR}).
As the first step we evaluate the integral
\begin{align}\label{Jzeta}
\mathfrak{R}(\zeta,z_1,z_2)=\int \frac{d^4x}{\pi^2}\frac{e^{-irx}}{x^2}\vev{p'|\mathcal{R}(z_1,z_2)|p}\,.
\end{align}
We take $r=q'+\zeta\Delta$ and will take the limit $\zeta\to z_1$ at the end of the calculation.
The relevant matrix elements of the ``kinematic'' twist-four operators
$\langle p'|\mathcal{O}_{1(2)}|p\rangle$ are written in terms of the leading-twist DD in Eq.~(\ref{ME:O12}).
Inserting these expressions in Eq.~(\ref{RbR}) and making use of the identity
\begin{equation}
(S_0-1)\frac{\ell_{ww'}^2}{w-w'}\int_0^1\! dv v\, e^{-iv(\ell_{ww'}x)}=
\frac{\ell_{ww'}^2}{w-w'}e^{-i(\ell_{ww'}x)}
\end{equation}
we obtain (note that we do not assume $z_{12}=1$ here)
\begin{eqnarray}
\vev{p'|\mathcal{R}(z_1,z_2)|p}&=&
-2i\int d\alpha d\beta\,\Phi(\beta,\alpha)
\int_{z_2}^{z_1}\!\!{dw}\int_{z_2}^{w}\!\!\frac{dw'}{z_{12}}
\nonumber\\
&&{}\hspace*{-0.7cm}\times\frac{w'-z_2}{z_1-w'}\biggl\{
\frac{\Delta^2}2S_+ +(S_0-1)(\Delta\ell_{ww'})
\nonumber\\
&&{}\hspace*{0.6cm}+\frac{i}{2}(\Delta x){\ell^2_{ww'}}
\biggr\}\frac{e^{-i(\ell_{ww'}x)}}{w-w'}\,.
\end{eqnarray}
Going over to the momentum space one gets after some algebra
\begin{eqnarray}
\mathfrak{R}(\zeta,z_1,z_2)&=&
\frac{2}{(n\tilde n)}\int d\alpha d\beta\,\Phi(\beta,\alpha)
\biggl[
\beta^2|P_\perp^2|\partial_F\hspace*{1.1cm}{}
\nonumber\\
&&{}\hspace*{-0.35cm}+\Delta^2(F\!-\!1)F\partial_F+\Delta^2\alpha
\biggr] I(\zeta,z_1,z_2)\,,
\end{eqnarray}
where
\begin{eqnarray}
\lefteqn{I(\zeta,z_1,z_2)=}
\\&=&
\int_{z_2}^{z_1}\frac{dw}{z_{12}}\int_{z_2}^{w}\frac{dw'}{z_1-w'}
\frac{w'-z_2}{\zeta-w-(w'-w) F-i\epsilon}
\nonumber\\&=&
\frac{1}{F-1}\int_{z_2}^{z_1}\frac{dw'}{z_{12}}
\frac{w'-z_2}{z_1-w'}
\ln\frac{\zeta-z_1-(w'-z_1)F-i\epsilon}{\zeta-w'-i\epsilon}\,.
\nonumber
\end{eqnarray}

The second and the third twist-four contributions in Eq.~(\ref{A++t234})
have similar structure, so that they can be added together.
These terms involve differential operators, $z_2\partial_2^2z_{12}$ and
$z_1\partial_1^2z_{21}$, acting on the field coordinates.
It is easy to show that
\begin{eqnarray}
z_2\partial_2^2z_{12}I(\zeta,z_1,z_2)&=&
\frac{1}{F-1}\frac{z_2}{z_{12}}\ln\frac{\zeta-z_1+z_{12}F-i\epsilon}{\zeta-z_2},
\nonumber\\
z_1\partial_1^2z_{21}I(\zeta,z_2,z_1)&=&-\frac{1}{F-1}\frac{z_1}{z_{12}}
\ln\frac{\zeta-z_2-z_{12}F-i\epsilon}{\zeta-z_1}.
\nonumber
\end{eqnarray}
Rescaling $z_i\to u z_i$ we can put $\zeta=z_1$ and $z_{12}=1$ and perform
the $u$-integration. The result reads
\begin{eqnarray}
 \mathcal{A}^{(2+3)}_{t=4}&=&
\frac{1}{2(n\bar n)}\int d\alpha d\beta\,\Phi(\beta,\alpha)
\nonumber\\&&{}\times
\Bigl[
\beta^2|P_\perp^2|\partial_F+\Delta^2(F-1)F\partial_F+\Delta^2\alpha
\Bigr]
\nonumber\\&\times&
\frac{1}{F-1}\biggl\{
{z_2}\Big[\frac{F}{F-z_1}\ln F-\frac{z_1}{z_2}\frac{F-1}{F-z_1}\ln z_1\Big]
\nonumber\\&&{}
+
{z_1}\Big[\frac{1-F}{z_2+F}\ln z_1-\frac{1-F}{z_2+F}\ln(1-F)\Big]
                   \biggr\},
\nonumber\\
\end{eqnarray}
where $F\to F-i\epsilon$ inside $z_2[\ldots]$  and $F\to F+i\epsilon$ inside $z_1[\ldots]$.

Finally, for the last twist-four contribution we need
\begin{eqnarray}
I(\zeta,z_2,z_1)|_{\zeta=z_1}&=& -\frac1F\ln(1-F-i\epsilon)
\nonumber\\&&{}\hspace*{-1.1cm}
+\frac1{F-1}\Big[
\Li_2(1)-\Li_2(F+i\epsilon)
\Big]\,,
\end{eqnarray}
where $\Li_2(1)= \pi^2/6$.
Changing variables in the first term, $\alpha\to -\alpha$, $\beta\to -\beta$, and using that under this
transformation $F\to 1-F$, one obtains
\begin{eqnarray}
\mathcal{A}^{(4)}_{t=4}&=&
-\frac{1}{(n\bar n)}\int d\alpha d\beta\,\Phi(\beta,\alpha)
\nonumber\\&&{}\times
\Bigl[
\beta^2|P_\perp^2|\partial_F+\Delta^2(F-1)F\partial_F+\Delta^2\alpha
\Bigr]
\nonumber\\&&{}\times
\frac1{F-1}\Big[\ln(F-i\epsilon)+\Li_2(F+i\epsilon)-\Li_2(1)
\Big].
\nonumber\\
\end{eqnarray}
Note that $\Li_2(x)$ is analytic at $x\to 0$ but has singularities in the
expansion around $x\to 1$, which is the reason that we keep the $+i\epsilon$ prescription
in $\Li_2(F+i\epsilon)$.

Collecting all twist-four terms and adding the contributions of twist-two and twist-three
operators we obtain the final result:
\begin{eqnarray}
A^{++}&=&\varkappa\int d\alpha d\beta\,\Phi(\beta,\alpha)\,\biggl\{2\log(F-i\epsilon)\hspace*{2.2cm}{}
\nonumber\\&& -\frac{1}{(n\tilde n)}\Big[\beta^2|P_\perp|^2\partial_F -\Delta^2(F-\alpha)\Big]\frac{1}{F-1}
\notag\\
&&{}\times
\!\Big[
\frac12\ln(F\!-\!i\epsilon)+\Li_2(F\!+\!i\epsilon)-\Li_2(1)
\Big]\!
\biggr\}.
\label{A++}
\end{eqnarray}
This expression is rather remarkable. First, notice that the dependence on $z_{1,2}$ disappeared
so that translation invariance is restored. Second, the result is finite: there are no infrared
divergences which might signal breakdown of factorization to the $1/Q^2$ accuracy because of a real photon
in the finite state, and would correspond to singularities in the limit $\zeta\to z_1$.
Third, singularities of the $1/Q^2$ correction at $F\to 0$ and $F\to 1$ (which correspond to
$x=\pm \xi$ in the GPD notation) are not stronger than the (logarithmic, in present variables)
singularity of the leading twist result. This property (see more details in the next
section) ensures that non-analyticity of the GPD at $x=\pm \xi$ does not lead to any problems,
there are no non-integrable singularities in the momentum fraction integral.

\section{Final expressions}\label{results}

In this Section we present our results for the helicity amplitudes in terms of the
GPD $H(x,\xi,t)$ (\ref{HPhi}).
Rewriting
$$
  F-\alpha = \frac{\beta}{\xi}+1-F
$$
we  left with integrals of the following structure:
\begin{eqnarray}
  I_1 &=& \int d\alpha d\beta\,\Phi(\beta,\alpha)\, Y(F)\,,
\nonumber\\
  I_2 &=& \int d\alpha d\beta\,\Phi(\beta,\alpha)\, \beta\, Y(F)\,,
\nonumber\\
  I_3 &=& \int d\alpha d\beta\,\Phi(\beta,\alpha)\, \beta^2\partial_F\, Y(F)\,,
\end{eqnarray}
where $Y(F)$ is a certain function of the $F$-variable defined in Eq.~(\ref{F}).
Inserting $1=\int_{-1}^1 dx\,\delta(x-\beta- \alpha\xi)$ under the integral
one obtains after a short calculation
\begin{eqnarray}
  I_1 &=& \int_{-1}^{1} dx\, Y\Big(\frac{x+\xi}{2\xi}\Big)\partial_x H(x,\xi,t)\,,
\end{eqnarray}
\begin{eqnarray}
  I_2 &=&
 \int_{-1}^{1} dx\,  Y\Big(\frac{x+\xi}{2\xi}\Big) (x\partial_x+\xi\partial_\xi) H(x,\xi,t)
\nonumber\\&=&
(\xi\partial_\xi-1) \int_{-1}^{1} dx\, Y\Big(\frac{x+\xi}{2\xi}\Big) H(x,\xi,t)\,,
\end{eqnarray}
and
\begin{eqnarray}
 I_3 &=& \!-2\xi\!\! \int_{-1}^{1}\!\!\! dx\,Y\Big(\frac{x\!+\!\xi}{2\xi}\Big) (\partial_xx\! +\!\xi\partial_\xi)
(x\partial_x\!+\!\xi\partial_\xi) H(x,\xi,t)
\nonumber\\&=&
-2\xi^3\partial_\xi^2 \int_{-1}^{1} dx\, Y\Big(\frac{x+\xi}{2\xi}\Big) H(x,\xi,t)\,,
\end{eqnarray}
where we used the identity
\begin{equation}
  \beta\,\delta(x-\beta- \alpha\xi)= (x\partial_x+\xi\partial_\xi)\theta(x-\beta- \alpha\xi)
\end{equation}
to arrive at the last two expressions.

Using these integrals it becomes straightforward to rewrite the results of the previous Section
in terms of $H(x,\xi,t)$. We obtain the helicity-flip amplitudes (in the notation of Eq.~(\ref{grading})
\begin{align}\label{A12a}
\mathcal{A}^{(1)}=&\frac{8\varkappa}{Q}
\,\xi^2\partial_\xi \int dx \frac{H(x,\xi,t)}{x-\xi}\ln\left(\frac{x+\xi}{2\xi}-i\epsilon\right)\,,
\notag\\[2mm]
\mathcal{A}^{(2)}=&\frac{8\varkappa}{Q^2}
\,\xi^3\partial_\xi^2
\int dx \frac{x\,H(x,\xi,t)}{x-\xi}\ln\left(\frac{x+\xi}{2\xi}-i\epsilon\right)
\end{align}
and the helicity-conserving amplitude
\begin{widetext}
\begin{align}\label{A0a}
\mathcal{A}^{(0)}=&-2\varkappa\Biggl\{\left(1-\frac{t}{2Q^2}\right)\int dx\frac{H(x,\xi,t)}{x+\xi-i\epsilon}
+ \frac{t}{Q^2}
\int dx\frac{H(x,\xi,t)}{x-\xi}\ln\left(\frac{x+\xi}{2\xi}-i\epsilon\right)
\notag\\
&
-\frac{2}{Q^2}\Big(\frac{t}{\xi}+2|P_\perp|^2\xi^2\partial_\xi\Big)\xi^2\partial_\xi
\int dx\frac{H(x,\xi,t)}{x-\xi}
\left[\frac12\ln\left(\frac{x+\xi}{2\xi}-i\epsilon
\right)
+\Li_2\left(\frac{x+\xi}{2\xi}+i\epsilon
\right)-\Li_2(1)
\right]
\Biggr\}\,.
\end{align}
\end{widetext}
The last expression presents the main result of this study.

The transverse momentum squared, $|P_\perp|^2 > 0$, can be expressed in terms of
kinematic invariants as (\ref{Pperp})
\begin{equation}
  |P_\perp|^2 = \frac14 (t_{\rm min}-t) \left(\frac1{{\xi^2}}-1\right),
\end{equation}
where
\begin{equation}
  t_{\rm min} = -4m^2\xi^2/(1-\xi^2)
\end{equation}
is the minimal kinematically allowed value of the momentum transfer $t$.
Note that the target mass corrections $\propto m^2$ only enter our results through the
dependence on $t_{\rm min}$ and are always overcompensated by the finite--$t$ corrections
$\propto |t|> |t_{\rm min}|$. By this reason target mass corrections alone do not have any
physical significance.

The definition of the skewedness parameter $\xi$ accepted in this work,
Eq.~(\ref{xi}), may not be the most natural one from the phenomenological point of view.
Let
\begin{equation}
   \xi_B = \frac{x_B}{2-x_B}\,,
\label{xiB}
\end{equation}
where $x_B = Q^2/(2pq)$ is the Bjorken scaling variable. A simple calculation yields for the relation
of ``our'' $\xi$ and $\xi_B$:
\begin{equation}
   \xi = \Big(1+\frac{t}{Q^2}\Big)\frac{\xi_B}{1+ \xi_B t/Q^2}\,.
\label{xi-xiB}
\end{equation}

We assume that the GPD $H(x,\xi,t)$ is continuous at the points $x=\pm\xi$ while its
first derivative can be singular. The renormalization-group analysis suggests that the
singularity in the derivative is at least logarithmic
$\partial_x H(x,\xi,t)\overset{x\to\pm\xi}{\sim} \ln (x\mp\xi)$
(if one does not employ some fine tuning).
It is easy to see that under this assumption the $x$-integrals in~Eqs.~(\ref{A12a}),(\ref{A0a}) converge
and define a smooth function of $\xi$ on the interval $(0,1)$. This statement is obvious for the
logarithmic terms,  whereas in order to see this for the dilogarithm terms (which have a branch point at $x=\xi$),
it is convenient to rewrite these contributions as follows:
\begin{eqnarray}
\lefteqn{\frac{1}{x-\xi}\left[\Li_2\Big(\frac{x+\xi}{2\xi}+i\epsilon\Big)-\Li_2(1)\right]=\hspace*{3cm}{}}
\nonumber\\&=&
\frac{1}{x-\xi}\int_0^{\frac{\xi-x}{2\xi}-i\epsilon}\! \frac{dt}{1-t}\ln
t\underset{x\to\xi}{\sim} \ln\Big(\frac{\xi-x}{2\xi}-i\epsilon\Big).~~~~
\end{eqnarray}

{}For completeness we also give here an explicit expression for the imaginary part of the
helicity-conserving amplitude:
\begin{eqnarray}
\lefteqn{\frac1{\pi}\text{Im} \mathcal{A}^{(0)}=}
\nonumber\\&=&2\varkappa\biggl\{\! H(\xi,\xi,t)\left(1\!-\!\frac{t}{2Q^2}\right)
\!+\!\frac{t}{Q^2}\!\int_{\xi}^{1}\!\!dx\frac{H(x,\xi,t)}{(x+\xi)}
\notag \\
&&{}+\frac{2}{Q^2}\Big(\frac{t}{\xi}+2|P_\perp|^2\xi^2\partial_\xi\Big)\xi^2\partial_\xi
\int^{1}_{\xi}\!\! dx{H(x,\xi,t)}
\notag\\
&&{}\times\Big[-\frac12\frac{1}{x+\xi}+\frac{1}{x-\xi}\ln\left(\frac{x+\xi}{2\xi}\right)\Big]\biggr\}\,,
\label{ImA++}
\end{eqnarray}
where we used that $H(-x,\xi,t)= -H(x,\xi,t)$ in order to bring the answer to this form.
Note that the imaginary part depends on the GPD in the DGLAP region only, $x>\xi$, in agreement
with dispersion relations.

\section{Numerical estimates and discussion}\label{numerics}

A detailed study of the numerical impact of kinematic power corrections goes beyond the tasks
of this paper. This study has to be done at the level of cross sections, 
taking into account finite--$t$ and target mass effects to kinematic (e.g. phase space) factors 
\cite{Belitsky:2010jw} and including
the interference with the Bethe--Heitler process. Since the corrections to helicity amplitudes
are calculated in this work for the simplified model case of a scalar target only, such a complete analysis is probably
not warranted at this stage.

In this Section we present numerical estimates for the kinematic power corrections to the imaginary part
of the helicity-conserving DVCS amplitude, Eq.~(\ref{ImA++}), which gives the largest contribution to the
cross section. To this end we use a model for the GPD $H(x,\xi, t)$ corresponding to the
$N=1$ ansatz from Ref.~\cite{Radyushkin:2011dh}.
It is based on the so-called single--DD description
which is defined  by the "gauge" fixing condition
$$
\alpha f(\beta,\alpha,t)=\beta g(\beta,\alpha, t)\,,
$$
imposed on the DDs $f$ and $g$ in Eq.~(\ref{DD1}),
see Ref.~\cite{Radyushkin:2011dh} for more details. It is assumed that the DD $f$ takes a
factorized form
\begin{align}
f(\beta,\alpha,t)=q(\beta,t) h(\beta,\alpha)\,.
\label{singleDD}
\end{align}
Here $q(x,t=0)$ is a (quark) parton distribution which we take as
\begin{equation}
q(x,t)=x^{-a(t)}(1-x)^3 e^{B t},
\end{equation} 
and
\begin{align}
h(\beta,\alpha)=\frac34\frac{(1-\beta)^2-\alpha^2}{(1-\beta)^3}.
\end{align}
The function $h(\beta,\alpha)$
satisfies the normalization condition $\int_{-1+|\beta|}^{1-|\beta|}d\alpha \,h(\beta,\alpha)=1$.
Note that we use $q(x)\sim (1-x)^3$ which is a standard approximation for the proton target,
because this is case that is interesting phenomenologically.
For the pion one expects $q(x)\sim (1-x)^{1-2}$.

We assume that the DD (\ref{singleDD}) depends on $t$ through the corresponding dependence of the Regge trajectory
$$
a(t)=0.48+0.9\, \text{GeV}^{-2} t
$$
and in addition involves a multiplicative factor $e^{B t}$ (see e.g. Ref.~\cite{Kroll:2010ad}).

The imaginary part of the amplitude (\ref{ImA++}) involves $H(x,\xi,t)$ in the region $x>\xi$ only.
In this region one obtains a compact expression~\cite{Radyushkin:2011dh}
\begin{align}
H(x,\xi,t)_{x>\xi}=\frac{3x}{4\xi}\int_{\beta_1}^{\beta_2}\frac{d\beta}{\beta^{1+a(t)}}
\biggl[\bar\beta^2-\left(\frac{x-\beta}{\xi}\right)^2\biggr] e^{Bt},
\end{align}
where $\beta_1=(x-\xi)/(1-\xi)$ and  $\beta_2=(x+\xi)/(1+\xi)$.

We consider the following ratios:
\begin{equation}
\frac{\text{Im}\mathcal{A}^{(0)}-\text{Im} \mathcal{A}_{LO}^{(0)}}{\text{Im} \mathcal{A}_{LO}^{(0)}}
= \frac{t}{Q^2} c_t(\xi,t)+ \frac{m^2}{Q^2} c_{m^2}(\xi,t)\,,
\end{equation}
where
\begin{equation}
  \text{Im} \mathcal{A}_{LO}^{(0)} = 2\varkappa\, H(\xi,\xi, t)
\end{equation}
is the leading-order leading-twist result.
Note that the $e^{B t}$ factor in the GPD cancels between the numerator and the denominator so that it is irrelevant
for our purposes.
The ratios $c_{t,m^2}(\xi,t)$ still depend on $t$, however, because of the non-factorizable $t$-dependence
of the GPD through the Regge trajectory.

\begin{figure}[t]
\psfrag{g}[cc][cc]{$c_t(\xi,t)$}
\psfrag{a}[cc][cc][0.9]{$t=0.0$}
\psfrag{b}[cc][cc][0.9]{$t=-0.2$}
\psfrag{c}[cc][cc][0.9]{$t=-0.5$}
\psfrag{d}[cc][cc][0.9]{$t=-1.0$}
\psfrag{e}[cc][cc][0.9]{$t=-1.5$}
\psfrag{x}[cc][cc]{$\xi$}
\includegraphics[width=8.5cm]{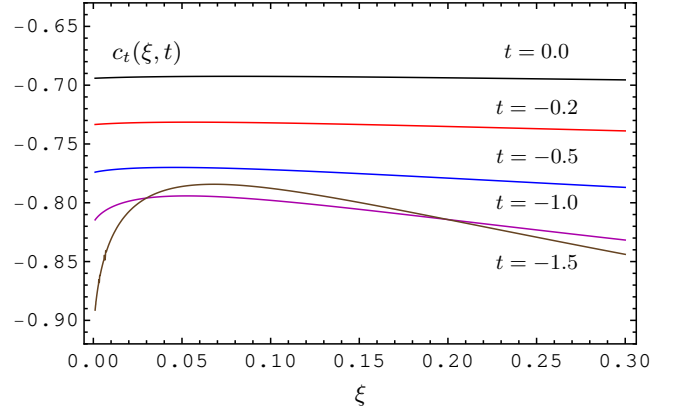}
\caption{The coefficient $c_t(\xi,t)$ in the range $0.001<\xi<0.3$ for  different values of $t$.
The curves correspond to
$-t=(0.0,\,0.2,\, 0.5,\, 1.0,\, 1.5)\,\text{GeV}^2$.}
\label{figure1}
\end{figure}

\begin{figure}[ht]
\psfrag{g}[cc][cc]{$c_{m^2}(\xi,t)$}
\psfrag{a}[bc][tc][0.9]{$t=0.0$}
\psfrag{b}[cc][cc][0.9]{$t=-0.2$}
\psfrag{c}[cc][cc][0.9]{$t=-0.5$}
\psfrag{d}[cc][cc][0.9]{$t=-1.0$}
\psfrag{e}[cc][cc][0.9]{$t=-1.5$}
\psfrag{x}[cc][cc]{$\xi$}

\includegraphics[width=8.5cm]{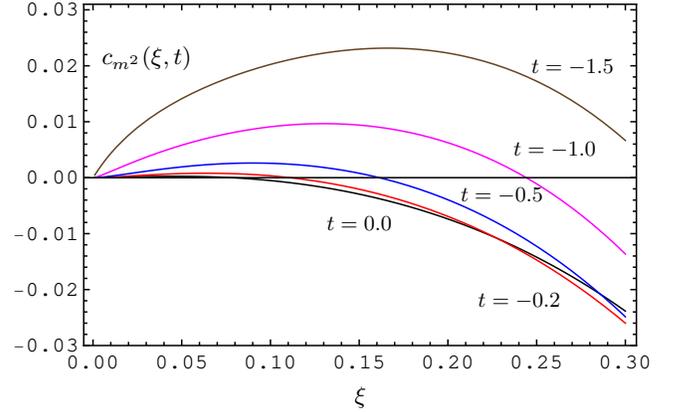}
\caption{The coefficient $c_{m^2}(\xi,t)$ in the range $0.001<\xi<0.3$ for  different values of $t$.
The curves  correspond  to $-t=(0.0,\,0.2,\, 0.5,\, 1.0,\, 1.5)\,\text{GeV}^2$.}
\label{figure2}
\end{figure}

In Fig.~\ref{figure1} we show $c_t(\xi,t)$ as a function of $\xi$ in the interval $0.001 < \xi < 0.3$
for $t\to 0$ (which also corresponds to a factorizable $t$ dependence of the GPD) and for
the values $-t=0.2, 0.5, 1.0$ and $1.5$~GeV$^2$. The same is shown in Fig.~\ref{figure2} for the coefficient $c_{m^2}(\xi,t)$.

One sees that in the whole kinematic range  $|c_t(\xi,t)|\gg |c_{m^2}(\xi,t)|$ so that the target mass correction to the imaginary
part of the amplitude is  almost negligible. The finite-$t$ correction is, on the other hand, quite sizable.
One has to have in mind that magnitude of the finite-$t$ correction depends on the definition of
the skewedness parameter. Using $\xi_B$ (\ref{xiB}) as a kinematic variable in the leading-order expression results in
a redefinition of the kinematic correction, which acquires a stronger $\xi$ dependence but does not become smaller.

For the particular case of the minimal allowed value of the momentum transfer $t=t_{\rm min}$ we find
\begin{equation}
\left.\frac{\text{Im}\mathcal{A}^{(0)}-\text{Im} \mathcal{A}_{LO}^{(0)}}{\text{Im} \mathcal{A}_{LO}^{(0)}}\right|_{t=t_{\rm min}}
\simeq  -(0.62- 0.65)
\frac{t_{\rm min}}{Q^2} \sim \frac{\xi^2m^2}{Q^2}
\end{equation}
which holds with a good accuracy in a broad interval of $\xi$ and $m^2$.

To summarize, in this paper we have presented the first complete
calculation of kinematic power corrections to the helicity amplitudes of
deeply-virtual Compton scattering to the twist-four accuracy
for a study case of a (pseudo)scalar target.
Our main result is that both finite--$t$, $\sim t/Q^2$, and
target mass, $\sim m^2/Q^2$, twist-four kinematic power corrections turn out to be
factorizable, at least to the leading order in the strong coupling.
Our numerical estimates using a certain simple  model of the generalized
parton distribution suggests that target mass contributions to helicity amplitudes
are very small, but the finite--$t$ effects, on the contrary, rather significant
in all kinematic regions of interest. The similar calculation for the DVCS from a nucleon
is in progress and the results will be published separately.

\section*{Acknowledgements}
V.B. is grateful to G.~Korchemsky and the Institute of Theoretical Physics at CEA Saclay for hospitality
and for the financial support, as well as the financial support by
the European Research Council under Advanced Investigator Grant ERC-AdG-228301
during his sabbatical stay at Saclay when this work was finalized.
This work was supported by the DFG, grant BR2021/5-2.

\appendix
\renewcommand{\theequation}{\Alph{section}.\arabic{equation}}
\section{Flavor--singlet operators}\label{App:A}
The generalization of the technique developed in Refs.~\cite{Braun:2011zr,Braun:2011dg} to the case of flavor-singlet
operators is straightforward, so that we will not go into technical details and discuss the key points only.
Our notation in this Appendix follows Ref.~\cite{Braun:2011dg}.

The flavor-singlet quark operator of the leading twist $O^q(z_1,z_2)\equiv{O}_{++}(z_1,z_2)$ defined in Eq.~(\ref{O++})
has positive parity under charge conjugation and gets mixed by the renormalization group equations with the twist-two gluon operator
\begin{align}
O^f(z_1,z_2)=&f^a_{++}(z_1n)\bar f^a_{++}(z_2n)\,.
\end{align}
The other existing quark-antiquark twist$-2$ operator
$ \sim O_V(z_1,z_2)+O_V(z_2,z_1)-O_A(z_1,z_2)+O_A(z_2,z_1)$ has negative $C$-parity.
It does not mix with gluon operators and it also does not contribute to the OPE.
The divergence of the flavor-singlet quark and gluon conformal operators
can be calculated following the procedure described in Ref.~\cite{Braun:2011dg}.
The result can be presented in the form of a scalar product
\begin{align}\label{E0}
(\partial\mathcal{O})^q_N=&\vev{\overrightarrow{\Psi}_1^q|\overrightarrow{\mathcal{Q}}}_q+
\vev{\overrightarrow{\Psi}_1^g|\overrightarrow{\mathcal{G}}}_g\,,
\notag\\
(\partial\mathcal{O})^f_N=&\vev{\overrightarrow{\Psi}_2^q|\overrightarrow{\mathcal{Q}}}_q+
\vev{\overrightarrow{\Psi}_2^g|\overrightarrow{\mathcal{G}}}_g\,,
\end{align}
where  $\overrightarrow{\mathcal{Q}}_q,\overrightarrow{\mathcal{G}}_g$ are the
flavor-singlet (anti)quark-gluon and three-gluon  {\it nonquasipartonic} operators, respectively,
and $\overrightarrow{\Psi}_{1,2}^q, \overrightarrow{\Psi}_{1,2}^g$ are the corresponding coefficient functions.
As discussed in detail in Ref.~\cite{Braun:2011dg}, the contributions of {\it quasipartonic}, alias four-particle
twist-four operators can be omitted.

The quark-gluon operators appearing in Eq.~(\ref{E0}) present a flavor-singlet analog of the
nonquasipartonic operators with ``good'' conformal properties introduced in~\cite{Braun:2011dg}:
\begin{align}
\mathcal{Q}_1(z)=&\Big[\bar\psi_+(z_1)t^a\psi_+(z_3)
+
\chi_+(z_3)t^a\bar\chi_+(z_1)\Big]f^a_{+-}(z_2)\,,
\notag\\
\mathcal{Q}_2(z)=&\Big[\bar\psi_+(z_1)t^a\psi_-(z_3)
+\chi_-(z_3)t^a\bar\chi_+(z_1)\Big]f^a_{++}(z_2)\,,
\notag\\
\mathcal{Q}_3(z)=&\Big[D_{-+}\bar\psi_+(z_1)t^a\psi_-(z_3)
\notag\\
&
+\chi_-(z_3)t^aD_{-+}\bar\chi_+(z_1)\Big]f^a_{++}(z_2)\,.
\label{eq:Q}
\end{align}
Explicit expressions for the three gluon operators, $\overrightarrow{\mathcal{G}}$, will not be relevant for
the further discussion.

The scalar product in (\ref{E0}) is determined by the requirement of
hermiticity of the evolution equations. Its existence is ensured
by conformal invariance.
We need to know an explicit form of the scalar product in the quark-gluon
sector only, and it turns out to be exactly the same as in the nonsinglet case.

It is clear that $(\partial\mathcal{O})^q_N$ does
not have a three-gluon component, so that $\overrightarrow{\Psi}_1^g=0$. The
wave functions~$\overrightarrow{\Psi}_1^q$ have the same form as in the nonsinglet case,
whereas $\overrightarrow{\Psi}_2^q$ require a new calculation. The results read:
\begin{eqnarray}
\Psi_1(w)&=&a_N\int_0^1d\alpha\,\alpha\bar\alpha\, (w_2-w_{13}^\alpha)^{N-1},
\notag\\
\Psi_2(w)&=&-\frac2{w_{12}}\partial_2w_{12}\partial_1 w_{12}^2\int_0^1d\alpha\,\bar\alpha\,
 (w_2-w^\alpha_{13})^{N-1},
\notag\\
\Psi_3(w)&=&b_N\int_0^1d\alpha\,\alpha\bar\alpha^2\, (w_2-w_{13}^\alpha)^{N-2},
\end{eqnarray}
where $\Psi_1(w) \equiv \Psi_1(w_1,w_2,w_3)$ etc.,
$a_N=-4\,(N+2)$ and $b_N=2\,(N-1)(N+2)(N+4)$.

Using orthogonality of the wave functions of different existing multiplicatively renormalizable operators
with respect to our scalar product, cf.~\cite{Braun:2011dg}, the contribution of the
divergence of the flavor-singlet twist-two operators $(\partial\mathcal{O})_N^{q,f}$
to the expansion of nonlocal operators $\overrightarrow{\mathcal{Q}},\overrightarrow{\mathcal{G}}$
can be found as
\begin{align}\label{EXP}
\begin{pmatrix}\overrightarrow{\mathcal{Q}}\\\overrightarrow{\mathcal{G}}\end{pmatrix}\,=\,&
\sum_N A_N\biggl\{\Big({||\Psi_2||^2 \Psi_1-\Psi_2\vev{\Psi_2|\Psi_1}}\Big)(\partial\mathcal{O})_N^{q}
\notag\\
&+\Big({||\Psi_1||^2 \Psi_2-\Psi_1\vev{\Psi_1|\Psi_2}}\Big)(\partial\mathcal{O})_N^{f}
\biggr\},
\end{align}
where $\Psi_k=(\overrightarrow{\Psi}_k^q,\overrightarrow{\Psi}_k^g)$,
$k=1,2$ and the coefficient $A_N$ is given by
\begin{equation}
A_N^{-1}=||\Psi_1||^2||\Psi_2||^2-|\vev{\Psi_1|\Psi_2}|^2.
\end{equation}
Note that in order to calculate $||\Psi_1||^2=||\Psi_1^q||_q^2$ and
$\vev{\Psi_1|\Psi_2}=\vev{\Psi_1^q|\Psi_2^q}_q$
it is sufficient to know the scalar product in the quark-gluon sector.

The three quark-gluon operators in Eq.~(\ref{eq:Q}) are not independent (see Ref.~~\cite{Braun:2011dg}) and their
contribution to the OPE of the time-ordered product of two electromagnetic currents can be expressed in terms of
$\mathcal{Q}_2$ alone. This contribution, in turn, can be written in terms of a certain integral over the gluon position
on the light cone:
\begin{align}
\int_{z_2}^{z_1}\!dw\, (w-z_2) \mathcal{Q}_2(z_1,w,z_2)\,.
\end{align}
Explicit calculation shows that
\begin{eqnarray*}
\int_{z_2}^{z_1}\!dw\, (w-z_2) (\overrightarrow{\Psi}_1^q)_2(z_1,w,z_2)&=&c_N||\Psi_1||^2\sim \gamma_N^{qq},
\nonumber\\
\int_{z_2}^{z_1}\!dw\, (w-z_2) (\overrightarrow{\Psi}^q_2)_2(z_1,w,z_2)&=&c_N\vev{\Psi_1|\Psi_2}\sim \gamma_N^{qg},
\end{eqnarray*}
where $\gamma_N^{qq}$ and $\gamma_N^{qg}$ are the corresponding entries in the matrix of anomalous dimensions
of the flavor-singlet twist-two operators.
It follows then from~(\ref{EXP}) that the gluon operator $(\partial\mathcal{O})_N^{f}$ does not
contribute to the OPE of two electromagnetic currents to our accuracy, while the flavor singlet quark operator
$(\partial\mathcal{O})_N^{q}$ enters the OPE with the same coefficient as in the nonsinglet case.

\section{Translation invariance of the $\mathcal{R}$--operator}\label{App:B}

The formal proof of translation invariance of our result~\cite{Braun:2011zr,Braun:2011dg}
for the T-product of two electromagnetic currents to the twist-four
accuracy, Eq.~(\ref{eq:shift2}), on the operator level is straightforward but rather long and technical.
In this paper we have chosen to write this result in terms of the $\mathcal{R}(z_1,z_2)$ operator defined in Eq.~(\ref{RbR})
which is translation-invariant itself, Eq.~(\ref{Rshift}). In this Appendix we give a proof of this relation,
which holds up to twist-six corrections.

Using the definition in Eq.~(\ref{RbR}) expression on the l.h.s. of~(\ref{Rshift})  can be
written in the form
\begin{equation}
z_{12}
\int_{z_2}^{z_1}\frac{d w}{z_{12}}\int_{z_2}^{w}\frac{d w'}{z_{12}} \frac{w'-z_2}{z_1-w'}\, Z(w,w')\,,
\end{equation}
where
\begin{eqnarray}
Z(w,w')&=&S_-\Big[\frac12S_+
 \mathcal{O}_{1}(w,w' )
-(S_0-1)\mathcal{O}_{2}(w,w')\Big]
\nonumber\\
&&{}\hspace*{-1.6cm}=
 \Big(\frac12S_+S_-+S_0\Big)\mathcal{O}_{1}(w,w' )
-S_0S_-\mathcal{O}_{2}(w,w')
\end{eqnarray}
and $S_-=\partial_w+\partial_{w'}$. The operators $\mathcal{O}_{1,2}$ are defined in
Eq.~(\ref{O12def}) and $\mathcal{O}^{t=2}_{++}$ in Eq.~(\ref{O++1}).
Taking into account the following identity for the leading-twist projector $\Pi$~(see Ref.~\cite{Braun:2011dg}):
\begin{equation}
\Pi(x,\lambda)\lambda^\alpha\bar\lambda^{\dot\alpha}=
\bar x^{\dot\alpha\alpha}\,\Pi(x,\lambda)-\frac12x^2\,\bar\partial^{\dot\alpha\alpha}\int_0^1\! du\, \Pi(ux,\lambda),
\end{equation}
one finds
\begin{eqnarray}
S_-\mathcal{O}_{++}^{t=2}(w,w')&=&\Pi(x,\lambda)[i(\mathbf{P}n),{O}_{++}(w,w')]
\nonumber\\&=&
{}[i(\mathbf{P}x),\mathcal{O}^{t=2}_{++}(w,w')]
\nonumber\\&&{}\hspace*{-0.3cm}
-\frac12x^2\!\int_0^1\!\!du u\,\mathcal{O}_2(uw,uw')\,.
\end{eqnarray}
With the help of this relation one easily obtains
\begin{eqnarray}
S_-\mathcal{O}_{1}(w,w')
&=&\big[i(\mathbf{P}x),\mathcal{O}_{1}(w,w')\big]+\ldots\,,
\notag\\
S_-\mathcal{O}_{2}(w,w')
&=&\big[i(\mathbf{P}x),\mathcal{O}_{2}(w,w')\big]+\mathcal{O}_{1}(w,w')
\notag\\
&&{}-\int_0^1 du u \big[i(\mathbf{P}x),\mathcal{O}_{2}(uw,uw')\big]
\nonumber\\&&{}+\ldots\,,
\end{eqnarray}
where the ellipses stand for the contributions of twist-six operators.
Finally, inserting these expressions in $Z(w,w')$ and taking into account that
\begin{equation}
S_0\int_0^1 du u \big[i(\mathbf{P}x),\mathcal{O}_{2}(uw,uw')\big]=
 \big[i(\mathbf{P}x),\mathcal{O}_{2}(w,w')\big]
\end{equation}
one easily verifies~Eq.(\ref{Rshift}).


\end{document}